\documentclass[12pt,a4paper]{article}

\RequirePackage{amsmath}
\RequirePackage{mathrsfs}
\RequirePackage{amsfonts}
\RequirePackage{dsfont}
\RequirePackage{braket}
\RequirePackage{tabularx}
\RequirePackage{nicefrac}
\RequirePackage{graphicx}
\RequirePackage{booktabs}
\RequirePackage{colortbl}
\RequirePackage{xcolor}
\RequirePackage[symbol]{footmisc}

\def\AEF{A.E. Faraggi}

\def\IJMP#1#2#3{{\it Int.\ J.\ Mod.\ Phys.}\/ {\bf A#1} (#2) #3}

\def\EJP#1#2#3{{\it Eur.\ Phys.\ Jour.}\/ {\bf C#1} (#2) #3}

\def\JHEP#1#2#3{{\it JHEP}\/ {\bf #1} (#2) #3}
\def\NPB#1#2#3{{\it Nucl.\ Phys.}\/ {\bf B#1} (#2) #3}
\def\PLB#1#2#3{{\it Phys.\ Lett.}\/ {\bf B#1} (#2) #3}
\def\PRD#1#2#3{{\it Phys.\ Rev.}\/ {\bf D#1} (#2) #3}
\def\PRL#1#2#3{{\it Phys.\ Rev.\ Lett.}\/ {\bf #1} (#2) #3}

\def\etal{{\it et al\/}}

\def\beq{\begin{equation}}
\def\eeq{\end{equation}}
\def\beqn{\begin{eqnarray}}
\def\eeqn{\end{eqnarray}}
\def\ds{{$\tilde S$}}
\def\unahe{{${\overline{\rm NAHE}}$}}

\newcommand{\CC}[2]{C{#1\atopwithdelims[]#2}}

\newcommand{\ba}{\begin{eqnarray}}
\newcommand{\ea}{\end{eqnarray}}

\begin{document}
\begin{titlepage}
\samepage{
\setcounter{page}{1}
\rightline{}
\rightline{November 2019}

\vfill
\begin{center}
{\Large \bf{
Stable Three Generation  Standard--like Model\\ \medskip
From a Tachyonic Ten Dimensional \\ \bigskip 
Heterotic--String Vacuum}}

\vspace{1cm}
\vfill

{\large Alon E. Faraggi\footnote{E-mail address: alon.faraggi@liv.ac.uk}, 
         Viktor G. Matyas\footnote{E-mail address: viktor.matyas@liv.ac.uk}
 and Benjamin Percival\footnote{E-mail address: benjamin.percival@liv.ac.uk} 
}
\\

\vspace{1cm}

{\it Dept.\ of Mathematical Sciences, University of Liverpool, Liverpool
L69 7ZL, UK\\}

\vspace{.025in}
\end{center}

\vfill
\begin{abstract}
\noindent
Recently it was proposed that the ten dimensional tachyonic 
superstring vacua may serve as good starting points for the 
construction of viable phenomenological models. 
Such phenomenologically viable models enlarge the space
of possible string solutions, and may offer novel insight 
into some of the outstanding problems in string phenomenology. 
In this paper we present a three generation standard--like model
that may be regarded as a compactification of a ten dimensional
tachyonic vacuum. We discuss the features of the model as compared 
to a similar model that may be regarded as compactification 
of the ten dimensional $SO(16)\times SO(16)$ heterotic--string. 
We further argue that in the four dimensional model all 
the geometrical moduli are fixed perturbatively, whereas the
dilaton may be fixed by hidden sector non--perturbative effects.

\end{abstract}

\smallskip}
\end{titlepage}

\section{Introduction}
\normalsize

The heterotic--string models in the free fermionic formulation \cite{fff}
provide a rich laboratory to develop the methodology of connecting string
theory with observational data. Since the late eighties this class 
of string compactifications produced an abundance of three generation
models 
\cite{fsu5,fny, alr, slm, lrs, acfkr, su62, frs, slmclass, lrsclass}, 
with qualitatively realistic properties, as well as an arena for 
investigation of cosmological scenarios \cite{coscos}. 
The relevant class of string 
compactifications are $Z_2\times Z_2$ orbifolds of six dimensional 
toroidal manifolds \cite{z2xz2}, 
that are related to compactifications on $Z_2$ 
orbifolds of $K_3\times T_2$ surfaces. This class of internal spaces
produces a rich symmetry structure also from a purely mathematical 
point of view \cite{tw}. 

Since the early days of string phenomenology, the majority of studies
have been devoted to the analysis of the $N=1$ supersymmetric string vacua.
Supersymmetry is then broken in the effective low energy field theory
limit by a gaugino or matter condensate. 
Electroweak radiative breaking also 
ensues by dimensional transmutation, and is compatible with the 
observed parameter space. 
However, while supersymmetry is an elegant theoretical construction,
the fact that it is not observed below the TeV scale lessens some 
of its motivating attributes. 
It is therefore vital to explore alternatives from the point
of view of string theory. 

Investigation of non--supersymmetric string models to date 
were conducted by studying compactifications of the tachyon--free 
$SO(16)\times SO(16)$ ten dimensional heterotic string theory 
\cite{dh, gv,  itoyama, kltclas, nonsusy, interpol, aafs}. 
This model can be generated as an orbifold of the ten
dimensional supersymmetric $E_8\times E_8$ heterotic--string, and 
the two vacua are connected by interpolation in a 
compactified dimension \cite{gv, itoyama}. Additionally, 
string theory gives rise to vacua
that are tachyonic in ten dimensions \cite{dh, gv, kltclas}.
Recently, it was proposed in ref. \cite{spwsp} that these ten dimensional 
string configurations may serve as viable starting points for 
constructing phenomenological models, 
and offer novel perspectives on some of the outstanding
issues in string phenomenology.
Tachyon--free 
four dimensional models may be constructed and ought to be taken on par 
with the non--supersymmetric $SO(16)\times SO(16)$ heterotic--string.
Moreover, they may 
reveal alternative symmetries to those provided by 
spacetime supersymmetry. An example is the Massive 
Boson--Fermion Degeneracy of \cite{msds}. 
It was demonstrated in ref. \cite{spwsp}
that the ten dimensional tachyonic modes
may be projected out by Generalised GSO projections. 
A standard--like tachyon free model was presented in ref. 
\cite{spwsp}, albeit with six generations rather than three. 
A tachyon--free three generation model in this class is still 
outstanding. 

In this paper, we present such a three generation Standard--like tachyon free 
Model that can be regarded as a compactification of a 
tachyonic ten dimensional string vacuum. 
We discuss the spectrum of the model and its distinct features compared 
to the supersymmetric and non--supersymmetric models emanating 
from the $E_8\times E_8$ and $SO(16)\times SO(16)$ heterotic--string models. 
Furthermore, we argue that the internal space in our construction is 
entirely fixed, which follows from the fact that all the untwisted 
geometrical moduli are projected out in the given model.
We suggest that in this model all the moduli, aside from the dilaton, 
are fixed perturbatively, 
whereas the dilaton may be fixed by a hidden sector racetrack 
mechanism \cite{racetrack}. As we elaborate in the discussion,
the internal structure of the model presented 
here is obtained from a previously constructed supersymmetric Standard--like
Model \cite{cfmt}. Some characteristics of the two models are consequently 
identical.  
Our present models might therefore be regarded as deformation of 
the supersymmetric model, which conforms with the conjecture in 
ref. \cite{spwsp} that all $(2,0)$ string compactifications are
connected via orbifolds or by interpolations. 

\section{Free fermionic constructions}\label{tendvacua}

In the free fermion formulation models are specified
in terms of boundary condition basis vectors and one--loop Generalised
GSO (GGSO) phases \cite{fff}. 
The $E_8\times E_8$ and $SO(16)\times SO(16)$ models in ten dimensions are
defined in terms of a common set of basis vectors 
\ba
v_1={\mathds{1}}&=&\{\psi^\mu,\
\chi^{1,\dots,6}| \overline{\eta}^{1,2,3},
\overline{\psi}^{1,\dots,5},\overline{\phi}^{1,\dots,8}\},\nonumber\\
v_{2}=z_1&=&\{\overline{\psi}^{1,\dots,5},
              \overline{\eta}^{1,2,3} \},\nonumber\\
v_{3}=z_2&=&\{\overline{\phi}^{1,\dots,8}\},
\label{tendbasisvectors}
\ea
where we adopted the conventional notation used in the
free fermionic constructions 
\cite{fsu5, slm, alr, lrs, acfkr, su62, frs, 
slmclass, lrsclass, su421, gkr, fknr, fkr, cfkr}. 
The basis vector $\mathds{1}$ is required by the consistency rules 
\cite{fff} and generates a model with an $SO(32)$ gauge
group from the Neveu-Schwarz (NS) sector. 
The spacetime supersymmetry generator 
is given by the combination 
\beq
S={\mathds{1}}+z_1+z_2 = \{{\psi^\mu},\chi^{1,\dots,6}\}. 
\label{tendsvector}
\eeq
The choice of Generalised GSO phase $\CC{z_1}{z_2}=\pm1$ selects between 
the $E_8\times E_8$ or $SO(16)\times SO(16)$ heterotic--strings 
in ten dimensions. The relation in eq. (\ref{tendsvector}) dictates that 
in ten dimensions the breaking of spacetime supersymmetry is correlated with 
the breaking pattern 
$E_8\times E_8\rightarrow SO(16)\times
SO(16)$. 
Equation (\ref{tendsvector}) does not hold in lower dimensions, 
and the two breakings are not correlated. 

It is noted that in both the $E_8\times E_8$ and $SO(16)\times SO(16)$ 
heterotic--strings in ten dimensions, the tachyonic states are projected out. 
The would--be tachyon in these models are obtained from the Neveu-Schwarz (NS)
sector, by acting on the right--moving vacuum with a single fermionic 
oscillator:
\beq
|~0\rangle_L\otimes {\bar\phi}^a|0\rangle_R
\label{untwistedtach}
\eeq
where in ten dimensions $a=1,\cdots, 32$. The GSO projection induced by the 
$S$--vector, which is the spacetime supersymmetry generator, projects 
out the untwisted tachyons, producing tachyon free models in both cases. 
As discussed in ref. \cite{spwsp} obtaining the 
ten dimensional tachyonic vacua in the free fermionic formulation amounts
to the removal of the $S$--vector from the construction. 
For example, the $SO(16)\times E_8$ heterotic--string model in ten dimensions
is generated by 
the basis vectors $\{{\mathds1}, z_1\}$ in eq. (\ref{tendbasisvectors}), 
independently of the GGSO phases. Other ten dimensional
configurations can similarly be obtained by substituting the $z_1$ basis vectors
with $z_1=\{{\bar\phi}^{1,\cdots,~4}\}$ and adding similar $z_i$ 
basis vectors, with four periodic fermions,
and utmost two overlapping.  
These vacua are connected by interpolations or orbifolds
along the lines of ref. \cite{gv}, and, in general, contain
tachyons in their spectrum. Our interest here is in the possibility of 
constructing tachyon free phenomenological vacua, starting from the 
tachyonic ten dimensional vacua.


As noted in the ten dimensional case, compactifications of the ten
dimensional tachyonic vacua amounts to removing the vector $S$
from the set of basis vectors. In four spacetime dimensions
the set $\{{\mathds1}, z_1,z_2\}$
produces a non supersymmetric model with $SU(2)^6\times SO(12)\times 
E_8\times E_8$ or $SU(2)^6\times SO(12)\times SO(16)\times SO(16)$. 
In this case the untwisted tachyonic states in general reappear.
It is noted also that the left--moving vector bosons remain in 
the spectrum, and are projected out by the additional NAHE--set
basis vectors. An alternative to removing the $S$--vector from 
the construction is to augment it with periodic right--moving 
fermions. A convenient choice is given by 
\beq
{\tilde S} = \{\psi^{1,2}, 
                \chi^{1,2},
                \chi^{3,4},
                \chi^{5,6}\vert {\bar\phi}^{1, \cdots,~4} \}\equiv~1~. 
\label{newS}
\eeq
In this case there are no massless gravitinos, and the untwisted 
tachyonic states 
\beq
|0\rangle_L\otimes {\bar\phi}^{1, \cdots,~4}|0\rangle_R 
\label{stildetachstates}
\eeq
are invariant under the ${\tilde S}$--vector projection.
We note that the untwisted tachyons are those that descend 
from the ten dimensional vacuum, hence confirming that the 
model can indeed be regarded as compactification of a 
ten dimensional tachyonic vacuum. 
The advantage of using the vector ${\tilde S}$ is that
its projection on the chiral generation is retained, hence 
facilitating the construction of three generation models. 
In ref. \cite{spwsp}, a three generation model with \ds~
was presented, which is, however, tachyonic.

Our tachyon free three generation model is constructed by using 
a modified NAHE--set \cite{nahe}, with the $S$--vector replaced 
by the \ds--vector. We refer to it as the ${\overline{\rm NAHE}}$--set.
The basis vectors of the ${\overline{\rm NAHE}}$--set are shown
in eq. (\ref{undernahe}). 

\beqn
 &&\begin{tabular}{c|c|ccc|c|ccc|c}
 ~ & $\psi^\mu$ & $\chi^{12}$ & $\chi^{34}$ & $\chi^{56}$ &
        $\bar{\psi}^{1,...,5} $ &
        $\bar{\eta}^1 $&
        $\bar{\eta}^2 $&
        $\bar{\eta}^3 $&
        $\bar{\phi}^{1,...,8} $ \\
\hline
\hline
      {\bf~1} & ~1 &~1&1&1 &~1,...,1 &~1 &~1 &~1 &~1~1~1~1~1~1~1~1 \\
          \ds & ~1 &~1&1&1 &~0,...,0 &~0 &~0 &~0 &~0~0~1~1~1~1~0~0 \\
\hline
  ${b}_1$ & ~1 &~1&0&0 &~1,...,1 &~1 &~0 &~0 &~0~0~0~0~0~0~0~0 \\
  ${b}_2$ & ~1 &~0&1&0 &~1,...,1 &~0 &~1 &~0 &~0~0~0~0~0~0~0~0 \\
  ${b}_3$ & ~1 &~0&0&1 &~1,...,1 &~0 &~0 &~1 &~0~0~0~0~0~0~0~0 \\
\end{tabular}
   \nonumber\\
   ~  &&  ~ \nonumber\\
   ~  &&  ~ \nonumber\\
     &&\begin{tabular}{c|cc|cc|cc}
 ~&      $y^{3,...,6}$  &
        $\bar{y}^{3,...,6}$  &
        $y^{1,2},\omega^{5,6}$  &
        $\bar{y}^{1,2},\bar{\omega}^{5,6}$  &
        $\omega^{1,...,4}$  &
        $\bar{\omega}^{1,...,4}$   \\
\hline
\hline
    {\bf~1} &~1,...,1 &~1,...,1 &~1,...,1 &~1,...,1 &~1,...,1 &~1,...,1 \\
    \ds     &~0,...,0 &~0,...,0 &~0,...,0 &~0,...,0 &~0,...,0 &~0,...,0 \\
\hline
${b}_1$ &~1,...,1 &~1,...,1 &~0,...,0 &~0,...,0 &~0,...,0 &~0,...,0 \\
${b}_2$ &~0,...,0 &~0,...,0 &~1,...,1 &~1,...,1 &~0,...,0 &~0,...,0 \\
${b}_3$ &~0,...,0 &~0,...,0 &~0,...,0 &~0,...,0 &~1,...,1 &~1,...,1 \\
\end{tabular}
\label{undernahe}
\eeqn
The model generated by eq. (\ref{undernahe}) produces some novel 
features. The untwisted tachyonic states are projected out by the 
projections of each of the basis vectors $b_i$ $i=1,2,3$. Hence, 
the model is tachyon free. For a suitable choice of 
GGSO phases, the four dimensional gauge group is 
$SO(10)\times SO(6)^3\times SO(16)$. Aside from the hidden
sector reduction $E_8\rightarrow SO(16)$, the gauge symmetry generated 
by (\ref{undernahe}) is identical to that generated by the NAHE--set. 
The novelty is in the structure of the chiral generations.
Whereas in models
that descend from the $SO(16)\times SO(16)$ heterotic--string 
the chiral generations may retain their supersymmetric structure,
up to some charges,  {\it i.e.}
the states from the sectors $S+b_i$ may remain massless and produce scalar 
states in the chiral $16$--representation of $SO(10)$ \cite{aafs},  
in \unahe~based models the states from the 
\ds$+b_i$--sectors do not produce massless states. 

The construction of three generation models in this class proceeds by 
adding three or four 
additional basis vectors to the \unahe--set. The additional
basis vectors break the $SO(10)$ GUT group to one of its subgroups, and 
simultaneously reduce the number of chiral generations to three generations. 
One from each of the sectors $b_1$, $b_2$ and $b_3$.
In addition to the spacetime vector bosons that produce the 
four dimensional gauge group, the untwisted sector produces
electroweak Higgs doublets that couple to the chiral generations from the 
sectors $b_i$ and can be used to generate fermion masses. In that respect 
\unahe--based models produce a structure which is similar to that of
NAHE--based models. The caveat is that in general the \unahe--based models 
will be tachyonic, which stems from the proliferation of tachyon
producing sectors, once the four dimensional gauge group is broken to 
smaller factors \cite{aafs}. 

Our three generation model is constructed as a variant of the Standard--like 
Model published in ref. \cite{cfmt}. The basis vectors that extend the 
\unahe--set and generate the Standard--like Model are given by 

\beqn
 &\begin{tabular}{c|c|ccc|c|ccc|c}
 ~ & $\psi^\mu$ & $\chi^{12}$ & $\chi^{34}$ & $\chi^{56}$ &
        $\bar{\psi}^{1,...,5} $ &
        $\bar{\eta}^1 $&
        $\bar{\eta}^2 $&
        $\bar{\eta}^3 $&
        $\bar{\phi}^{1,...,8} $ \\
\hline
\hline
  ${\alpha}$  & ~0 &~0&0&0 &~1~1~1~0~0 &~1 &~0 &~0 &~1~1~0~0~0~0~0~0 \\
  ${\beta}$   & ~0 &~0&0&0 &~1~1~1~0~0 &~0 &~1 &~0 &~0~0~1~1~0~0~0~0 \\
  ${\gamma}$  & ~0 &~0&0&0 &
		${1\over2}$~${1\over2}$~${1\over2}$~${1\over2}$~${1\over2}$
	      & ${1\over2}$ & ${1\over2}$ & ${1\over2}$ &
               ~0~0~0~0~$1\over2$~$1\over2$~${1\over2}$~${1\over2}$ \\
\end{tabular}
   \nonumber\\
   ~  &  ~ \nonumber\\
   ~  &  ~ \nonumber\\
     &\begin{tabular}{c|c|c|c}
 ~&   $y^3{y}^6$
      $y^4{\bar y}^4$
      $y^5{\bar y}^5$
      ${\bar y}^3{\bar y}^6$
  &   $y^1{\omega}^5$
      $y^2{\bar y}^2$
      $\omega^6{\bar\omega}^6$
      ${\bar y}^1{\bar\omega}^5$
  &   $\omega^2{\omega}^4$
      $\omega^1{\bar\omega}^1$
      $\omega^3{\bar\omega}^3$
      ${\bar\omega}^2{\bar\omega}^4$ \\
\hline
\hline
$\alpha$ &~1 ~~~~0 ~~~~0 ~~~~1  &~0 ~~~~0 ~~~~1 ~~~~1  &~0 ~~~~0 ~~~~1 ~~~~1 \\
$\beta$  &~0 ~~~~0 ~~~~1 ~~~~1  &~1 ~~~~0 ~~~~0 ~~~~1  &~0 ~~~~1 ~~~~0 ~~~~1 \\
$\gamma$ &~0 ~~~~1 ~~~~0 ~~~~0  &~0 ~~~~1 ~~~~0 ~~~~0  &~1 ~~~~0 ~~~~0 ~~~~0 \\
\end{tabular}
\label{stringmodel}
\eeqn
The basis vectors $\alpha$, $\beta$ and $\gamma$ are identical to those used
in \cite{cfmt}. Modular invariance constraints necessitates that the 
GGSO phases are modified. However, the only modifications are
in the phases that involve the phases associated with the 
basis vector ${\tilde S}$, 
with the choice of generalised GSO coefficients:

\begin{equation}
{\bordermatrix{
        &{\bf~1}&  \tilde{S} & &{b_1}&{b_2}&{b_3}& &{\alpha}&{\beta}&{\gamma}\cr
 {\bf~1}&   ~~1 &~~1 & & -1  &  -1 & -1  & &  -1    &  -1   & ~~i   \cr \tilde{S}&   ~~1 &~~1 & &~~1  & ~~1 &~~1  & &  -1    &  -1   & ~~i   \cr
        &       &    & &     &     &     & &        &       &       \cr
   {b_1}&    -1 & -1 & & -1  &  -1 & -1  & &  -1    &  -1   & ~~i   \cr
   {b_2}&    -1 & -1 & & -1  &  -1 & -1  & &  -1    & ~~1   & ~~i   \cr
   {b_3}&    -1 & -1 & & -1  &  -1 & -1  & & ~~1    &  -1   & ~~1   \cr
	&       &    & &     &     &     & &        &       &       \cr
{\alpha}&    -1 & -1 & & -1  &  -1 &~~1  & & ~~1    & ~~1   & ~~1   \cr
 {\beta}&    -1 &~~1 & & -1  & ~~1 & -1  & &  -1    & ~~1   & ~~1   \cr
{\gamma}&    -1 & -1 & &~~1  & ~~1 & -1  & &  -1    &  -1   &  -i   \cr}}
\label{phasesmodel1}
\end{equation}

In some respects therefore the vacuum defined by eqs. (\ref{undernahe},
\ref{stringmodel}) 
and (\ref{phasesmodel1}) shares some of the properties of the model
of ref. \cite{cfmt}. These similarities are particularly noted with respect to 
the untwisted sector and the sectors $b_1$, $b_2$ and $b_3$ 
that produce the Standard Model spectrum. The two vacua are
of course entirely different as the one in ref. \cite{cfmt} is 
supersymmetric, whereas the one defined by eqs. (\ref{undernahe},
\ref{stringmodel},
\ref{phasesmodel1}) is not. We further remark that the model 
of ref. \cite{cfmt} can be used to explore tachyon free compactifications of
$SO(16)\times SO(16)$ heterotic string, similar to the models studied
in \cite{aafs}. This is obtained by using the basis vectors of 
ref. \cite{cfmt},
{\it i.e.} with an unmodified $S$--vector, but with the change of phases
\begin{equation} 
\CC{S}{\beta}=\CC{S}{\gamma}=-1~\rightarrow~\CC{S}{\beta}=\CC{S}{\gamma}=+1.
\label{phasechanges}
\end{equation}
The resulting spectrum is tachyon free. With these modifications the 
only sectors that break supersymmetry are sectors that extend the NAHE--set. 
Hence, in this case the chiral matter spectrum still exhibits
a supersymmetry--like structure, as discussed in \cite{aafs}. 

Turning back to the model defined by (\ref{undernahe}, \ref{stringmodel}) and 
(\ref{phasesmodel1}), as discussed in ref. \cite{cfmt}, 
the basis vectors of the model utilise both symmetric and asymmetric 
boundary conditions with respect to the sets of internal 
worldsheet fermions 
$\{y\vert{\bar y}\}^{3,\cdots,6}$, 
$\{y^{1,2}, \omega^{1,2}\vert{\bar y}^{1,2},{\bar\omega}^{5,6}\}$, and
$\{\omega\vert{\bar\omega}\}^{1,\cdots,4}$. Each of 
these three sets is periodic 
in $b_1$, $b_2$ and $b_3$, respectively. 
This assignment induces the doublet--triplet splitting mechanism \cite{dtsm}
on the untwisted $5+{\bar5}$ multiplets, where symmetric assignment
keeps the triplets and projects the doublets, and vice versa for the
asymmetric assignment. The novelty in the model of ref. \cite{cfmt},
and as can be seen from eq. (\ref{stringmodel}) 
is that both symmetric and asymmetric boundary conditions are utilised
with respect to the two sectors $b_1$ and $b_2$, whereas only 
asymmetric boundary conditions are utilised with respect to $b_3$. 
The result is that both the untwisted doublet and triplets, {\it i.e.} the 
entire $5+{\bar 5}$ representations are projected from the first 
and second planes that couple to the states from the sectors $b_1$ and 
$b_2$, whereas the third untwisted plane produces a pair of electroweak 
Higgs doublets that couples to the states from the sector $b_3$ at 
leading order.
This model, like the model of ref. \cite{cfmt}, contains one additional
scalar Higgs doublet beyond the Standard Model.
We note that the weak doublet scalar states
$H_{2},~{\bar H}_2, ~
H_{46},~{\bar H}_{46}, ~
H_{47},~{\bar H}_{47}, ~
H_{56},~{\bar H}_{56}, ~
H_{57},~{\bar H}_{57}$
in tables \ref{exotics1}--\ref{nonSmapped4},
are exotic vector--like states that should receive a
heavy mass. Such states are common in free fermionic models. The
electroweak doublet states
${\tilde h}, ~\bar{\tilde{h}}~
H_{8},~{\bar H}_{8}, ~
H_{13},~{\bar H}_{13} $
in tables \ref{stildetab} and \ref{exotics1} are
spacetime fermionic states. Hence, the only available
electroweak Higgs doublets are those arising in the NS--sector.
A leading top quark Yukawa coupling is obtained at the cubic
level of the potential due to the boundary condition assignment
in the $\gamma$--basis vector \cite{topyuk}. 
%
An unintended consequence of these assignments is 
that the untwisted moduli space is reduced substantially, due to the 
projection of additional $SO(10)$ singlet fields from the spectrum. 
This led to the suggestion in ref. \cite{cfmt} that all the moduli
in the model, aside from the dilaton, are fixed. This structure is 
expected to persist in the present model. 

Spacetime vector bosons in the 
model defined by (\ref{undernahe}, 
\ref{stringmodel}) and (\ref{phasesmodel1}) 
are obtained only from the untwisted sector. The observable gauge symmetry
is defined by the charges carried by the observable chiral matter states
{\it i.e.} those arising from the sectors $b_1$, $b_2$ and $b_3$. It is 
given by: 
\beq
SU(3)_C\times SU(2)_L\times U(1)_C\times U(1)_L\times U(1)_{1,2,3}\times
U(1)_{4,5,6}~~~ .
\label{observablegg}
\eeq
where, 
\begin{eqnarray}
U(1)_C& = & {\rm Tr}\, U(3)_C~\Rightarrow~Q_C=
			 \sum_{i=1}^3Q({\bar\psi}^i)~,\label{u1c}\\
U(1)_L& = & {\rm Tr}\, U(2)_L~\Rightarrow~Q_L=
			 \sum_{i=4}^5Q({\bar\psi}^i)~.\label{u1l}
\end{eqnarray}
The flavour $U(1)_{1,2,3}$ are generated by 
the worldsheet complex fermions ${\bar\eta}^{1,2,3}$ 
whereas $U(1)_{4,5,6}$ are generated by ${\bar\zeta}^{1,2,3}$. 
The complex fermions ${\bar\zeta}^i$ are defined as
${\bar\zeta}^1=(1/\sqrt{2})({\bar y}^3+{\bar y}^6)$,
${\bar\zeta}^2=(1/\sqrt{2})({\bar y}^1+{\bar\omega}^5)$ and 
${\bar\zeta}^3=(1/\sqrt{2})({\bar\omega}^2+{\bar\omega}^4)$. 
Each of the sectors $b_1$, $b_2$ and $b_3$ is charged with respect 
to $U(1)_i$ and $U(1)_{i+3}$. The appearance of the additional 
$U(1)_{4,5,6}$ symmetries arises due to the use of asymmetric 
boundary conditions that are essential for fixing the 
geometrical moduli \cite{moduli}. We note that this structure 
of the observable gauge symmetries is similar to that of the 
Standard--like Models in \cite{slm}. 

The hidden sector of the model arises from the complex worldsheet fermions 
${\bar\phi}^{1,\cdots,8}$ and is given by
\beq
SU(2)_{1,\cdots,~6}\times U(1)_{7,8}, 
\label{hiddengg}
\eeq
where $U(1)_{7,8}$ symmetries correspond to the combinations for 
worldsheet charges 
\begin{equation}
Q_{7}=\sum_{i=5}^6Q({\bar\phi}^i)~{\rm and}~
Q_{8}=\sum_{i=7}^8Q({\bar\phi}^i).\label{qh1}
\end{equation}
In NAHE--based models, the vector combination 
\beq
\zeta={\mathds1}+b_1+b_2+b_3,
\label{zetasec}
\eeq
may give additional spacetime vector bosons that enhance the hidden sector
gauge gauge, which are, however, projected out in the model defined 
by eqs. (\ref{undernahe},\ref{stringmodel}) and (\ref{phasesmodel1}), 
and the hidden sector is not enhanced.
The retention/projection of the enhancing states from the $\zeta$--sector
correspond to the $x$--map of ref. \cite{xmap}.
The hidden sector gauge group differs from that of ref. \cite{cfmt}
due to the right--moving periodic fermions in the ${\tilde S}$--vector. 
The Neveu--Schwarz sector produces in addition to the graviton, 
dilaton, antisymmetric tensor and spacetime vector bosons, 
one pair of electroweak Higgs doublets ${h}^3$ and ${\bar h}^3$; 
six pairs of $SO(10)$ singlet fields, which are charged with respect
to $U(1)_{4,5,6}$; and three fields that are neutral under the entire 
four dimensional gauge. These NS scalar fields are the same as those
that are obtained in the model of ref. \cite{cfmt}.
The two model differ in the fermionic spectrum generated in the 
$S$-- and ${\tilde S}$--sectors, respectively,
and in any combination that contains these vectors, 
on which we elaborate below.
The full massless spectrum of the model is presented in 
Appendix A. All sectors, fermonic and bosonic, have CPT conjugates 
that are not displayed explicitly in the tables in Appendix A. 

\subsection{Analysis of the Spectrum}
As mentioned, the model under investigation takes the model of 
\cite{cfmt} and transforms $S$ to $\tilde{S}$ and applies the 
phase changes in eq.  (\ref{phasechanges}). The states in the Hilbert 
space of this model are presented in Appendix A. It is worth 
exploring further the action of $\tilde{S}$-map on sectors in this 
model and how it differs from the SUSY generator, 
which is induced by the $S$--map.
In supersymmetric vacua the superpartners of the states from a given sector 
$\rho$ are obtained from the sector $S+\rho$. In non--supersymmetric models 
in which supersymmetry is broken by a GGSO phase \cite{aafs}, 
the states from the sector $S+\rho$ may be projected out, but more
generally they remain in the Hilbert space with modified charges. Hence,
these sectors retain the Fermi--Bose degeneracy of the massless 
states. Additionally, in such models, in general, there are sectors for which
the states in the sector $S+\rho$ are massive. These sectors 
therefore do not preserve the Fermi--Bose degeneracy. 
Additionally, the $\zeta$--sector induces the $\zeta$--map. 
In models with enhanced hidden sector gauge group, the $\zeta$--mapped
sectors $\rho+\zeta$, complements the states from a sector $\rho$ 
to representations of the enhanced symmetry group. In models in which 
the enhancing states are projected out, the states will be mapped 
to other representations. However, in many cases the 
number of states is preserved overall. This phenomenon
was observed {\it e.g.} in the case of spinor--vector duality 
of ref. \cite{fkr,cfkr,ffmt} under the $x$--map and reflects 
the modular properties of the underlying partition function. 
Similar properties may therefore be exhibited in the models with 
broken supersymmetry
that reflect their underlying modular properties. 
Due to the assignment of Ramond boundary conditions to 
$\bar{\phi}^{3456}$ in $\tilde{S}$ there are many sectors such as 
those listed in Tables \ref{nonSmapped2}, \ref{nonSmapped1},
\ref{NonSmapped3} and \ref{nonSmapped4} for this model, where no 
$\tilde{S}$-mapped states appear in the spectrum. The most 
common reason for this is that the $\tilde{S}$-mapped sector gains 
additional contributions to the mass formula on the right due to the 
addition of the Ramond $\bar{\phi}^{3456}$ in $\tilde{S}$, thus 
making the $\tilde{S}$-mapped sector massive. 
The most notable sectors of this type are the $b_i$ $i=1,2,3$ sectors
that produce the three chiral generations, one from each of the sectors 
$b_1$, $b_2$ and $b_3$, whereas the bosonic states in the sectors 
${\tilde S}+b_i$ $i=1,2,3$ are massive. This should be contrasted 
with the model generated by employing the phase modification in 
eq. (\ref{phasechanges}), which breaks supersymmetry, but retains 
the scalar states from the sectors $S+b_i$ $i=1,2,3$.
However, the sector $1+\tilde{S}+b_3+\alpha+\beta$  shown in table 
\ref{hiddentable}, is projected out by the GGSO 
projections from the \unahe-basis, despite giving rise to massless 
states \textit{a priori}. In NAHE-based models such a projection 
of a super--partnered state cannot occur.
We also note the existence of fermionic states in the ${\tilde S}$--sector
that transform as doublets of the electroweak symmetry and an hidden 
$SU(2)$ gauge group suggesting the possible implementation of electroweak
symmetry breaking by fermion condensates. 

The sectors shown in table \ref{so10sinws} exhibit the mapping 
property mentioned above. The left--moving sector of the ${\tilde S}$--vector
is the same as that of the $S$--vector. Hence, 
the ${\tilde S}$--vector still maps 
spacetime fermions to spacetime bosons· The $2\gamma$--map correspond 
to the $x$--map of ref. \cite{xmap}, which is a map between the 
spinorial sectors $b_j$, $j=1,2,3$, to the vectorial producing sectors 
$b_j+2\gamma$ (or $b_j+x$). The $\zeta$--map then correspond to 
the map to sectors that supplement the hidden sector representations
when the hidden sector gauge symmetry is enhanced, and otherwise 
maps to other states that descend from the massive spectrum. The states
in these sectors are therefore arranged in three groups of four.
Understanding the detail structure of the spectrum is crucial not only for 
understanding the properties of a single model, but rather in order 
to understand the global structure that underlies the larger space
of vacua, as demonstrated by the spinor--vector duality 
\cite{fkr,cfkr,ffmt}. 
Another observation of the $\tilde{S}$-map is that for certain 
sectors the addition of the Ramond $\bar{\phi}^{3456}$ can 
change the mass formula on the right so as to map a spinorial 
sector to a vectorial one. Such an outcome is observed, 
for example, in sector $b_3\pm \gamma$ which is spinorial, 
whereas $\tilde{S}+b_3\pm \gamma$ is vectorial.

As is common in $(2,0)$ heterotic--string 
compactifications \cite{cfua1}, the model generated by the basis vectors in 
(\ref{undernahe}, \ref{stringmodel}) and GGSO phases in
(\ref{phasesmodel1}) contains six $U(1)$ symmetries with 
non--vanishing trace, 
\beqn
{\rm Tr}U(1)_1 ={\rm Tr}U(1)_2 ={\rm Tr}U(1)_3 = -24 \nonumber\\
{\rm Tr}U(1)_4 ={\rm Tr}U(1)_5 ={\rm Tr}U(1)_6 = -12. \label{tru1tou6}
\eeqn
Five combinations of these $U(1)$'s are anomaly free and one combination, 
given by 
\beq
U(1)_A= 2 (U(1)_1+U(1)_2+U(1)_3) + (U(1)_4+U(1)_5+U(1)_6),
\label{u1ua}
 \eeq
remains anomalous. 
The anomalous $U(1)$
is removed by the Green--Schwarz--Dine--Seiberg--Witten mechanism
\cite{gs,dsw},
but gives rise to a tadpole diagram in string perturbation theory
at one--loop order  \cite{ads},
which reflects the instability of the string vacuum.
The mismatch between the bosonic and fermionic states at different
mass levels produces a non--vanishing vacuum energy, which
similarly gives rise to a tadpole diagram, indicating the
instability of the string vacuum. We may contemplate the possibility
of suppressing one against the other so that they conspire to
cancel. The anomalous $U(1)$ contribution is proportional
to the trace over the massless fermionic states
and the sign can be altered by the GGSO projections \cite{ads,cfua1}.
Both the anomalous $U(1)$ and the vacuum energy will depend on further
details of the model, which are very complicated, {\it e.g.} the 
potential of the remaining scalar fields in the spectrum, that can
shift the vacuum.
A comprehensive analysis is beyond our scope here, and possibly out
of reach in terms of the contemporary tools available due to the large number
of fields in our quasi--realistic model. However, we note that the 
same issues also plague the supersymmetric vacua with an anomalous 
$U(1)$ as well as vacua that are compactifications of the
non--supersymmetric $SO(16)\times SO(16)$ heterotic--string.
Therefore, a shift of the vacuum is either legitimate, or illegitimate,
in both cases. Any statement about the stability of the string vacua 
is at best speculative. We therefore propose that all of the 
non--tachyonic string vacua should be considered on equal footing. 
We can compactify the different ten dimensional vacua on the same 
internal structure and try to learn from the properties at the 
different limits. In the next section we illustrate this 
with regard to the question of stability of the model.


\section{Moduli Fixing}
Next we discuss the question of the moduli in the string model
defined by eqs. (\ref{stringmodel}) and (\ref{phasesmodel1}). 
The issue of moduli in string compactification is intricate. It is only
properly understood in compactifications with $(2,2)$ worldsheet supersymmetry
and with at least $N=1$ spacetime supersymmetry \cite{dkl}, 
but not in the more generic $(2,0)$ compactifications. 
Nevertheless, we can borrow from the terminology in those cases. 
The fact that 
supersymmetric and non--supersymmetric string vacua can be interpolated 
\cite{itoyama, interpol}, suggest that the moduli in the more generic 
compactifications with $(2,0)$ worldsheet supersymmetry can be
interpolated to the fields in the corresponding $(2,2)$ compactifications
\cite{spwsp}. To study the moduli in the model defined by eq. 
(\ref{stringmodel}) and (\ref{phasesmodel1}) we follow the discussion
in ref. \cite{moduli}. The geometrical moduli in the model are 
identified in terms of worldsheet Thirring interactions \cite{thirring}
that are invariant under the fermionic transformation properties 
defined by a given set of basis vectors, and are parameterised
by untwisted fields in the massless string spectrum \cite{moduli}.
For symmetric orbifold models, the exactly marginal operators
associated with the untwisted moduli fields have the general form
$\partial X^I{\bar\partial} X^J$, where $X^I$, $I=1,\cdots,6$, are
the coordinates of the six--torus $T^6$. 
The untwisted
moduli fields in this models admit the geometrical interpretation of
background fields, which appear as couplings of
the exactly marginal operators in the non--linear sigma model action.
The untwisted moduli scalar fields are the background fields that are 
compatible with the orbifold symmetry. 
In the fermionic 
formalism the exactly marginal operators are given in terms of Abelian
Thirring operators of the form $J_L^i(z){\bar J}_R^j({\bar z})$,
where $J_L^i(z)$, ${\bar J}_R^j({\bar z})$
are some left-- and right--moving $U(1)$ chiral currents
described by worldsheet fermions. 
The untwisted moduli correspond to the Abelian Thirring
interactions that are compatible with the GGSO projections
induced by the boundary condition basis vectors, 
in a given string model. 

The set of Abelian Thirring operators, and
untwisted moduli fields, is restricted by the projections 
induced by progressive boundary condition basis vectors. 
The minimal basis set in the model defined by eqs. 
(\ref{stringmodel}, \ref{phasesmodel1}) contains the
two vectors $\{{\mathds1}, {\tilde S}\}$. 
This set generates a non--supersymmetric tachyonic model 
with $SO(36)\times SO(8)$ right--moving gauge group. The
tachyonic states are the surviving untwisted tachyonic states 
in eq. (\ref{stildetachstates}). 
As in the ten dimensional tachyon free vacua, the six
$\chi_I$ are identified with the fermionic superpartners
of the six bosonic coordinates. This is because the 
${\tilde S}$--vector preserves the left--moving structure
of the $S$--vector in the corresponding non--tachyonic vacua. 
Each pair $\{y^i, \omega^i\}$ is identified with the fermionised 
version of the corresponding left--moving bosonic coordinate $X^i$, 
{\it i.e.} $i\partial X_L^i\sim y^i\omega^i$. The two dimensional 
action of the Abelian worldsheet Thirring interactions is
\beq
S~=~\int d^2z h_{ij}(X) J_L^i(z) {\bar J}_R^j({\bar z})~,
\label{2daction}
\eeq
where $J_L^i (i=1,\cdots ,6)$ are the left--moving chiral currents of
$U(1)^6$ 
and ${\bar J}^j_R (j=1,\cdots,22)$, are 
the right--moving chiral currents $U(1)^{22}$. 
The couplings $h_{ij}(X)$, as
functions of the spacetime coordinates $X^\mu$, are four 
dimensional scalar fields that are identified with the 
untwisted moduli fields. In the model
with the two basis vectors $\{{\mathds1}, {\tilde S}\}$
the $6\times22$ fields $h_{ij}(X)$ in eq. (\ref{2daction})
correspond to the 21 and 15 components
of the background metric $G_{IJ}$ and antisymmetric tensor
$B_{IJ}$ $(I,J=1,\cdots,6)$, plus the $6\times 16$ Wilson lines
$A_{Ia}$. The $h_{ij}(X)$ fields parameterise the
$SO(6,22)/SO(6)\times SO(22)$ coset--space of the toroidally
compactified space. The $h_{ij}$ untwisted moduli fields arise from
the NS sector, 
\beq
\vert \chi^I\rangle_L ~\otimes \vert{\bar\Phi}^{+J}{\bar\Phi}^{-J}\rangle_R~,
\label{untwisted22moduli}
\eeq
given in terms of the 22 complex right--handed world--sheet fermions
${\bar\Phi}^{+J}$ and their complex conjugates
${\bar\Phi}^{-J}$. The corresponding
marginal operators are given as
\beq
J^i_L(z){\bar J}^j_R({\bar z}) ~\equiv~~:y^i(z)\omega^i(z)(z):
:{\bar\Phi}^{+j}({\bar z}){\bar\Phi}^{-j}({\bar z}):~.
\label{marginal22operators}
\eeq
It is seen that the transformation properties of $\chi^i$, which appear
in the moduli (\ref{untwisted22moduli}), are the same as those of 
$y^i\omega^i$, which appear in the Abelian Thirring interactions
(\ref{marginal22operators}). We further note that the supersymmetric
vacuum defined by $\{{\mathds1}, S\}$ and the non--supersymmetric 
vacuum defined by  $\{{\mathds1}, {\tilde S}\}$ can be connected 
by continuous interpolations, by turning on the appropriate Wilson
lines, as demonstrated in the corresponding ten dimensional cases 
\cite{gv}. The important observation is that the modification
of the basis vector $S\rightarrow {\tilde S}$ does not affect the 
untwisted scalar moduli space which is therefore identical in the
two vacua, as well as in the corresponding non--supersymmetric 
model that descends from the $SO(16)\times SO(16)$ heterotic--string 
in ten dimensions. 

The ensuing analysis of the untwisted moduli follows closely that of 
refs. \cite{moduli} and \cite{cfmt}.
Adding the basis vectors $b_1$, $b_2$ and $b_3$, reduces the 
space of untwisted moduli scalars to the three sets 
\beq
h_{ij}=\vert \chi^i\rangle_L \otimes 
\vert{\bar y}^j{\bar\omega}^j\rangle_R=
\begin{cases}
(i,j=1,2)& $ $\cr
(i,j=3,4)& $ $\cr
(i,j=5,6)& $ $,\cr
\end{cases}
\label{hij}
\eeq
that parameterise the moduli space
\beq
{\cal M}=\left({{SO(2,2)}\over {SO(2)\times SO(2)}}\right)^3.
\eeq
These untwisted moduli fields are present in all symmetric 
$Z_2\times Z_2$ orbifolds. In complexified form they correspond to 
three K\"ahler and three complex structures of the $Z_2\times Z_2$ 
orbifold \cite{moduli}. As in the case of the model of \cite{cfmt}
The addition of the three basis vectors
beyond the \unahe--set results in the projection of all the 
states in eq. (\ref{hij}), {\it i.e.} all of the geometrical moduli 
are projected out. One can further check that the scalar states
arising from the NS--sector are indeed identical in the two models.
It should be emphasised that this outcome is particular to the 
boundary condition assignment for the set of left--moving 
real fermions $\{ y , \omega\}^{1,\cdots,~6}$ and their specific pairings
\cite{moduli}. The basic result is that due to this particular assignment
all the internal circles of the six dimensional torus are shifted 
asymmetrically, hence fixing the moduli of all six circles simultaneously, 
which is possible only in the case of the $Z_2\times Z_2$ orbifold.

$Z_2\times Z_2$ orbifold models also contain moduli from the twisted sectors.
It was argued in \cite{moduli} that these moduli are also projected 
out from the massless spectrum in our string model. 
In the supersymmetric vacua this follows from the reduction
of $E_8\times E_8 \rightarrow SO(16)\times S0(16)$
by the basis vector set 
$\{ {\mathds1}, S, \zeta_1={\mathds1}+b_1+b_2+b_3,~2\gamma\}$.
To identify the twisted moduli in the fermionic $Z_2\times 
Z_2$ orbifolds, it is instrumental to consider the set 
$\{ {\mathds1}, S, \zeta_1, x\}$, with 
$x=\{{\bar\psi}^{1,\cdots, 5}, {\bar\eta}^{1,2,3}\}$ \cite{xmap}. 
The $Z_2\times Z_2$ orbifold in the $E_8\times E_8$ case
reduces the observable $E_8$ symmetry to $E_6\times U(1)^2$, 
and produces states in the $27$ representation of $E_6$ from the 
twisted sectors. Under the decomposition of $E_6\rightarrow SO(10)\times
U(1)$, the $27$ multiplet split as $16_{1/2}+10_{-1}+1_2$, in a convenient 
normalisation of the $U(1)$ generator. In the free fermionic construction
the $16$ spinorial components are obtained from the sectors 
$b_i$, whereas the $10+1$ components are obtained from the sectors
$b_i+x$. The sectors $b_i+x$ produce an additional $E_6$ singlet field, 
which is identified with the twisted moduli \cite{xmap, moduli}. 
The class of vacua with enhanced $E_6$ symmetry possess $(2,2)$ 
worldsheet supersymmetry. In vacua in which the $E_6$ symmetry
is reduced to $SO(10)\times U(1)$, the $(2,2)$ worldsheet supersymmetry
is reduced to $(2,0)$. The states from the sectors 
$b_i+x$ are mapped to hidden sector matter states \cite{xmap, moduli}, 
{\it i.e.} the twisted moduli are projected out. The only states
that arise from the twisted sectors in this case are the 
observable and hidden sector matter states. These states have superpotential 
mass terms and therefore should not be identified as moduli. 
It should be emphasised, 
though, that any discussion of the twisted moduli in the $(2,0)$ vacua 
should be taken with a grain of salt, as their proper identification is 
only possible in vacua with $(2,2)$ worldsheet supersymmetry 
\cite{dkl, stefan}. 

More to the point, however, is the analysis of supersymmetric flat directions
that was carried out in ref. \cite{cfmt}. It was observed there that a
certain class of flat directions, which are designated as ``stringent flat 
directions'', do not exist in the model analysed in \cite{cfmt}. It was 
further argued that ``stringent flat directions'' are the only flat 
directions that are exact to any order in the superpotential, and that 
non--stringent flat directions must be broken at some order. In that case, 
it was argued that all the moduli in the model are fixed, where the geometrical
moduli are fixed by the asymmetric boundary conditions, whereas the 
supersymmetric moduli are fixed by the absence of exact supersymmetric 
flat directions, and the dilaton may be fixed by hidden sector 
non--perturbative effects \cite{cfmt}. It was further argued that 
supersymmetry is broken pertubatively in the model due to the
existence of an anomalous $U(1)$ symmetry that produces
a Fayet--Iliopolous term at one--loop \cite{dsw}. 

As discussed above, in respect to geometrical moduli, the supersymmetric 
model of \cite{cfmt} and the non--supersymmetric model discussed here, are 
identical, as the map $S\rightarrow {\tilde S}$ does not affect the 
geometrical moduli. Borrowing from the discussion of the absence of 
flat directions in the supersymmetric case, we argue that also in the 
present vacuum all the moduli are fixed. The argument is that the 
internal space in the two vacua is identical and is not affected
by the map $S\rightarrow {\tilde S}$, as can be seen from the assignment
of the remaining boundary condition basis vectors in eq. (\ref{stringmodel}) 
and in ref. \cite{cfmt}.

We note here that in general in non--supersymmetric string vacua
one expects that at a certain order in perturbation theory
manifest supersymmetry breaking is also communicated to the
scalar potential, lifting any flat directions. However,
there are several caveats to this expectation. In the first place,
the models are connected by deformations to points in the moduli
space that admit tachyonic states. It is therefore
not entirely clear that a given non--supersymmetric string
model can stabilise at a finite value of the moduli. In the same
vein, stabilisation of the moduli is a dynamical problems
involving a large number of scalar fields. Whether or not
all the moduli can stabilise at a finite value is a very
hard problem that occupies much of the discourse in string phenomenology
over the past two decades, and is still on going.
Our argument here does not rely on such considerations.
In our string model the geometrical moduli are simply projected
out. The model is therefore by construction not connected
to any points in the moduli space that admits tachyonic states.
Similarly, the geometric moduli are not merely stabilised.
They are frozen. Our model therefore provides an example of
non--supersymmetric string vacua in which all the moduli, aside
from the dilaton, are fixed, irrespective of the dynamics of
vacuum stabilisation. Whether the vacuum itself is stable hinges
on the possibility of stabilising the dilaton at finite
value, and with positive vacuum energy. We alluded here
to the possibility of using the racetrack mechanism \cite{racetrack}
to stabilise the dilaton. The question of the vacuum
energy in the class pertaining models will be reported
in a future publication. However, we should stress that
we regard our model as exploratory, providing some
insight into novel possibilities in string phenomenology, rather
than aiming to address the full set of questions involved.
Even if one could demonstrate that a non--supersymmetric
stable string model with suppressed positive vacuum energy exists
at one--loop, its viability at higher order in string perturbation
theory will still be open, unlike the case for supersymmetric vacua. 

\section{Discussion and Conclusion}\label{conclusion}

In this paper we presented a tachyon free 
three generation standard--like model
that may be regarded as a compactification of a 
tachyonic ten dimensional vacuum.
The model is non--supersymmetric and tachyon--free. 
It represents a new class of phenomenological string vacua,
with notable differences compared to vacua that can be 
built on the same internal structure. In this example, we considered
the supersymmetric compactification of ref. \cite{cfmt}
as well as the construction of a model in which supersymmetry 
is broken by  GGSO phase, {\'a} la ref. \cite{aafs}.
This non--supersymmetric version can be regarded as descending
from the non--supersymmetric $SO(16)\times SO(16)$ heterotic---string 
in ten dimensions.  
Both this non--supersymmetric vacuum, as well as the supersymmetric
model of ref. \cite{cfmt},
utilise the same set of boundary condition basis vectors 
with the substitution ${\tilde S}\rightarrow S$. 
In all three cases the untwisted NS scalar spectrum is the same, indicating
that the internal structure in all three models is identical.
The twisted spectrum is, however, entirely different. In the model
presented here the states from the sectors ${\tilde S} + b_j$ are 
massive, whereas in the other non--supersymetric model the corresponding
states from the sectors $S+b_j$ remain in the massless spectrum, albeit
with modified $U(1)$ charges. Hence in this case the chiral generations
still exhibit a supersymmetric like structure, although they do not reside 
in super--multiplets. These examples illustrate how compactifications
of the different ten dimensional vacua can be used to explore 
the phenomenological properties on the same internal structure. 
In that respect compactifications of the tachyonic ten dimensional vacua
may reveal new insight into outstanding issues in string phenomenology. 

Furthermore, as the internal structure of the model was adopted 
from the previously constructed supersymmetric model in ref. \cite{cfmt}, 
we used the observation made there in regard to the absence of stringent 
flat direction, and consequently the fixing of all moduli, to argue that
in our present model all moduli, aside from the dilaton, are also fixed 
perturbatively, whereas the dilaton may be fixed by hidden sector 
non--perturbative dynamics. In ref. \cite{cfmt} it was argued that the
absence of exact flat directions suggests that supersymmetry is broken
perturbatively in the model due to the existence a Fayet--Iliopoulos 
term at one--loop. In our present model supersymmetry is broken explicitly
at the Planck scale, but we carried forward the argument from ref. 
\cite{cfmt} to propose that all the moduli in the current model 
are also fixed and hence the vacuum would be stable. We should warn, however, 
that any discussion of stability in non--supersymmetric string vacua is
speculative, fraught with uncertainty, and possibly premature. 
Nevertheless, we note that the conditions that enable us 
to speculate on this stability are very particular to the configuration 
exhibited in this particular class of standard--like models. For example 
it was observed in ref. \cite{cleaveretal} that in flipped 
$SU(5)$ string vacua with internal structure similar to the one
used in ref. \cite{cfmt}, there do exist stringent supersymmetric
flat directions. The absence of stringent flat directions is therefore
specific to vacua with similar internal structure and standard--like model
gauge group. Similarly, as noted in ref. \cite{moduli}
the projection of the untwisted 
geometrical moduli is specific to the pairing of the internal
worldsheet fermions employed in our model. In this regard, we note that
while the standard--like model in ref. \cite{fny} shares many of the
phenomenological characteristics with the standard--like model of ref. 
\cite{slm}, whereas in the later case the geometrical moduli are fixed, 
in the former case they are not. It is seen again that this property
is specific to a particular configuration and is not generic. We may
infer conservatively that the stability issue of the string vacuum
can only be addressed in the very specific string vacua that come
close to describing the Standard Model, rather than in the generic 
cases.

\section*{Acknowledgments}

AEF would like to thank the Weizmann Institute for hospitality,
where part of this work was conducted. The work of VGM is supported in
part by EPSRC grant EP/R513271/1. The work of BP is supported in 
part by STFC grant ST/N504130/1. 

\bigskip
\newpage

\begin{appendix}
\section{The Spectrum of the Model}
The following tables present the spectrum of the model given in Section \ref{tendvacua}.
All charges are multiplied by four and the CPT conjugates are omitted for all states.
Throughout the tables we will make use of the vector combination:
$\zeta=1+b_1+b_2+b_3=\{\bar{\phi}^{1,...,8}\}$.\\

\small{
\begin{table}[h]
\setlength{\tabcolsep}{2.25pt}
\renewcommand{\arraystretch}{1}
  \begin{tabular}{| c | c | c | c | c c c c c c c c | c | c c |}
    \hline

    F & Sector & Name & $(C,L)$ & $Q_C$ & $Q_L$ & $Q_{\bar{\eta}^1}$ & $Q_{\bar{\eta}^2}$ & $Q_{\bar{\eta}^3}$ & $Q_{\bar{y}^{3,6}}$ & $Q_{\bar{y}^1\bar{w}^5}$ & $Q_{\bar{w}^{2,4}}$ & $SU(2)_{1,...,6}$ &  $Q_{7}$ & $Q_{8}$ \\ \hline

     b & NS & ($h$)                  & (1,2)   &~0 & -4 &~0 &~0 &~4  &~0  &~0  &~0  & (1,1,1,1,1,1) & ~0 &~0 \\
       &    & ($\bar{h}$)          & (1,$2$) &~0 &~4  &~0 &~0 & -4 &~0  &~0  &~0  & (1,1,1,1,1,1) & ~0 &~0 \\
       &    & $(\Phi_{56})$        & (1,1)   &~0 &~0  &~0 &~0 &~0  &~4  &~4  &~0  & (1,1,1,1,1,1) & ~0 &~0 \\
       &    & $(\bar{\Phi}_{56})$  & (1,1)   &~0 &~0  &~0 &~0 &~0  & -4 & -4 &~0  & (1,1,1,1,1,1) & ~0 &~0 \\
       &    & $(\Phi_{56}')$       & (1,1)   &~0 &~0  &~0 &~0 &~0  & -4 &~4  &~0  & (1,1,1,1,1,1) & ~0 &~0 \\
       &    & $(\bar{\Phi}_{56}')$ & (1,1)   &~0 &~0  &~0 &~0 &~0  &~4  & -4 &~0  & (1,1,1,1,1,1) & ~0 &~0 \\
       &    & $(\Phi_{46})$        & (1,1)   &~0 &~0  &~0 &~0 &~0  &~4  &~0  &~4  & (1,1,1,1,1,1) & ~0 &~0 \\
       &    & $(\bar{\Phi}_{46})$  & (1,1)   &~0 &~0  &~0 &~0 &~0  & -4 &~0  & -4 & (1,1,1,1,1,1) & ~0 &~0 \\
       &    & $(\Phi_{46}')$       & (1,1)   &~0 &~0  &~0 &~0 &~0  & -4 &~0  &~4  & (1,1,1,1,1,1) & ~0 &~0 \\
       &    & $(\bar{\Phi}_{46}')$ & (1,1)   &~0 &~0  &~0 &~0 &~0  &~4  &~0  & -4 & (1,1,1,1,1,1) & ~0 &~0 \\
       &    & $(\Phi_{45})$        & (1,1)   &~0 &~0  &~0 &~0 &~0  &~0  &~4  &~4  & (1,1,1,1,1,1) & ~0 &~0 \\
       &    & $(\bar{\Phi}_{45})$  & (1,1)   &~0 &~0  &~0 &~0 &~0  &~0  & -4 & -4 & (1,1,1,1,1,1) & ~0 &~0 \\
       &    & $(\Phi_{45}')$       & (1,1)   &~0 &~0  &~0 &~0 &~0  &~0  & -4 &~4  & (1,1,1,1,1,1) & ~0 &~0 \\
       &    & $(\bar{\Phi}_{45}')$ & (1,1)   &~0 &~0  &~0 &~0 &~0  &~0  &~4  & -4 & (1,1,1,1,1,1) & ~0 &~0 \\
       &    & $(\xi_{1,2,3})$      & (1,1)   &~0 &~0  &~0 &~0 &~0  &~0  &~0  &~0  & (1,1,1,1,1,1) & ~0 &~0 \\ \hline
  \end{tabular}
  \caption{The untwisted Neveu-Schwarz scalar states. All charges are multipled by 4.}
\end{table}
}
\small{
\begin{table}[h]
\setlength{\tabcolsep}{2.25pt}
\renewcommand{\arraystretch}{1}
  \begin{tabular}{| c | c | c | c | c c c c c c c c | c | c c |}
    \hline

    F & Sector & Name & $(C,L)$ & $Q_C$ & $Q_L$ & $Q_{\bar{\eta}^1}$ & $Q_{\bar{\eta}^2}$ & $Q_{\bar{\eta}^3}$ & $Q_{\bar{y}^{3,6}}$ & $Q_{\bar{y}^1\bar{w}^5}$ & $Q_{\bar{w}^{2,4}}$ & $SU(2)_{1,...,6}$ &  $Q_{7}$ & $Q_{8}$ \\ \hline
     f & $\tilde{S}$ & $\tilde{h}$& (1,2) &~0 &~4  &~0 &~0 &~0  &~0 &~0 &~0 &  (1,1,1,2,1,1) & -4 &~0 \\
       &     &$\bar{\tilde{h}}$ & (1,$2$) &~0 & -4 &~0 &~0 &~0  &~0 &~0 &~0 &  (1,1,1,2,1,1) &~4  &~0 \\
       &     &$\xi_4$ & (1,1)         &~0 &~0  &~0 &~0 &~0  &~0 &~0 &~0 &  (1,1,2,1,2,1) &~0  &~0 \\
       &     &$\xi_5$ & (1,1)         &~0 &~0  &~0 &~0 &~0  &~0 &~0 &~0 &  (1,1,2,1,2,1) &~0  &~0 \\
       &     &$\phi_1$ & (1,1)         &~0 &~0  &~0 &~0 &~4  &~0 &~0 &~0 &  (1,1,1,2,1,1) & -4 &~0 \\
       &     &$\phi_2$ & (1,1)         &~0 &~0  &~0 &~0 & -4 &~0 &~0 &~0 &  (1,1,1,2,1,1) &~4  &~0 \\ \hline
    f & $\tilde{S}+\zeta$ &$\xi_6$ &(1,1) &~0 &~0 &~0  &~0 &~0 &~0 &~0 &~0 & (1,2,1,1,2,1) & ~4 & -4 \\
      &                   &$\xi_7$& (1,1) &~0 &~0 &~0  &~0 &~0 &~0 &~0 &~0 & (1,2,1,1,2,1) & -4 & ~4 \\
      &                   &$\phi_3$&(1,1) &~0 &~0 & ~4 &~0 &~0 &~0 &~0 &~0 & (2,1,1,1,1,1) & ~0 & -4 \\
      &                   &$\phi_4$ &(1,1) &~0 &~0 & -4 &~0 &~0 &~0 &~0 &~0 & (2,1,1,1,1,1) & ~0 & ~4 \\
    \hline
  \end{tabular}
  \caption{The $\tilde{S}$ and $\tilde{S}+\xi$ sectors. All charges are multipled by 4 and the combination $\zeta=1+b_1+b_2+b_3$ has been used.}
  \label{stildetab}
\end{table}
}

\small{
\begin{table}[h]
\setlength{\tabcolsep}{2.25pt}
\renewcommand{\arraystretch}{1}
  \begin{tabular}{| c | c | c | c | c c c c c c c c | c | c c |}
    \hline

    F & Sector & Name & $(C,L)$ & $Q_C$ & $Q_L$ & $Q_{\bar{\eta}^1}$ & $Q_{\bar{\eta}^2}$ & $Q_{\bar{\eta}^3}$ & $Q_{\bar{y}^{3,6}}$ & $Q_{\bar{y}^1\bar{w}^5}$ & $Q_{\bar{w}^{2,4}}$ & $SU(2)_{1,...,6}$ &  $Q_{7}$ & $Q_{8}$ \\ \hline

    f & $b_1$ & $Q_1$    & (3,2)         &~2  &~0  & -2 &~0 &~0 & -2 &~0 &~0 & (1,1,1,1,1,1) &~0 &~0 \\
      &       & $u_1^c$  & ($\bar{3}$,1) & -2 & -4 & -2 &~0 &~0 &~2  &~0 &~0 & (1,1,1,1,1,1) &~0 &~0 \\
      &       & $d_1^c$  & ($\bar{3}$,1) & -2 &~4  & -2 &~0 &~0 & -2 &~0 &~0 & (1,1,1,1,1,1) &~0 &~0 \\
      &       & $L_1$    & (1,2)         & -6 &~0  & -2 &~0 &~0 &~2  &~0 &~0 & (1,1,1,1,1,1) &~0 &~0 \\ 
      &       & $e_1^c$  & (1,1)         &~6  &~4  & -2 &~0 &~0 &~2  &~0 &~0 & (1,1,1,1,1,1) &~0 &~0 \\
      &       & $N_1^c$  & (1,1)         &~6  & -4 & -2 &~0 &~0 & -2 &~0 &~0 & (1,1,1,1,1,1) &~0 &~0 \\\hline

    f & $b_2$ & $Q_2$    & (3,2)         &~2  &~0  &~0 & -2 &~0 &~0 &~2  &~0 & (1,1,1,1,1,1) &~0 &~0 \\
      &       & $u_2^c$  & ($\bar{3}$,1) & -2 & -4 &~0 & -2 &~0 &~0 & -2 &~0 & (1,1,1,1,1,1) &~0 &~0 \\
      &       & $d_2^c$  & ($\bar{3}$,1) & -2 &~4  &~0 & -2 &~0 &~0 &~2  &~0 & (1,1,1,1,1,1) &~0 &~0 \\
      &       & $L_2$    & (1,2)         & -6 &~0  &~0 & -2 &~0 &~0 & -2 &~0 & (1,1,1,1,1,1) &~0 &~0 \\ 
      &       & $e_2^c$  & (1,1)         &~6  &~4  &~0 & -2 &~0 &~0 & -2 &~0 & (1,1,1,1,1,1) &~0 &~0 \\
      
      &       & $N_2^c$  & (1,1)         &~6  & -4 &~0 & -2 &~0 &~0 &~2  &~0 & (1,1,1,1,1,1) &~0 &~0 \\\hline

    f & $b_3$ & $Q_3$   & (3,2)         &~2  &~0  &~0 &~0 & ~2 &~0 &~0 & -2 & (1,1,1,1,1,1) &~0 &~0 \\ 
      &       & $u_3^c$ & ($\bar{3}$,1)         & -2 & -4 &~0 &~0 & ~2 &~0 &~0 &~2  & (1,1,1,1,1,1) &~0    &~0 \\
      &       & $d_3^c$ & ($\bar{3}$,$2$) & -2 &~4  &~0 &~0 & ~2 &~0 &~0 & -2 & (1,1,1,1,1,1) &~0 &~0 \\
      &       & $L_3$   & (1,2)         & -6 &~0  &~0 &~0 & ~2 &~0 &~0 &~2  & (1,1,1,1,1,1) &~0 &~0 \\
      &       & $e_3^c$ & ($1$,$1$)     &~6  &~4  &~0 &~0 & ~2 &~0 &~0 &~2  & (1,1,1,1,1,1) &~0 &~0 \\
      &       & $N_3^c$ & (1,1)         &~6  & -4 &~0 &~0 & ~2 &~0 &~0 & -2 & (1,1,1,1,1,1) &~0 &~0 \\
     \hline

  \end{tabular}
  \caption{The observable matter sectors. All charges are multipled by 4.}
\label{nonSmapped2}
\end{table}
}
\small{
\begin{table}[h]
\setlength{\tabcolsep}{2.25pt}
\renewcommand{\arraystretch}{1}
  \begin{tabular}{| c | c | c | c | c c c c c c c c | c | c c |}
    \hline

    F & Sector & Name&$(C,L)$ & $Q_C$ & $Q_L$ & $Q_{\bar{\eta}^1}$ & $Q_{\bar{\eta}^2}$ & $Q_{\bar{\eta}^3}$ & $Q_{\bar{y}^{3,6}}$ & $Q_{\bar{y}^1\bar{w}^5}$ & $Q_{\bar{w}^{2,4}}$ & $SU(2)_{1,...,6}$ & $Q_{7}$ & $Q_{8}$ \\ \hline

    b & $b_1 + b_2$      &$\Phi^{\alpha\beta}_1$& (1,1) &~0 &~0 &~0 &~0 &~0 & ~2 & ~2 &~0 & (1,2,1,2,1,1) &~0 &~0 \\
      & $+ \alpha+\beta$ &$\bar{\Phi}^{\alpha\beta}_1$& (1,1) &~0 &~0 &~0 &~0 &~0 & -2 & -2 &~0 & (1,2,1,2,1,1) &~0 &~0 \\
      &                  &$\Phi^{\alpha\beta}_2$& (1,1) &~0 &~0 &~0 &~0 &~0 & -2 & ~2 &~0 & (2,1,2,1,1,1) &~0 &~0 \\
      &                  &$\bar{\Phi}^{\alpha\beta}_2$& (1,1) &~0 &~0 &~0 &~0 &~0 & ~2 & -2 &~0 & (2,1,2,1,1,1) &~0 &~0 \\ \hline

    f & $\tilde{S}+b_1+b_2$ &$\tilde{\Phi}^{\alpha\beta}_3$& (1,1) &~0 &~0 &~0 &~0 &~0 &~2  & -2 &~0 & (2,1,2,1,1,1) &~0 &~0 \\
      & $+\alpha+\beta $    &$\bar{\tilde{\Phi}}^{\alpha\beta}_3$& (1,1) &~0 &~0 &~0 &~0 &~0 & -2 &~2  &~0 & (2,1,2,1,1,1) &~0 &~0 \\ \hline

    f & $\tilde{S}+b_1+b_2$ &$\Phi^{\alpha\beta}_4$& (1,1) &~0 &~0 &~0 &~0 &~0 & ~2 & -2 &~0 & (1,1,1,2,1,2) &~0 &~0 \\
      & $+\alpha +\beta+\zeta$  &$\bar{\Phi}^{\alpha\beta}_4$& (1,1) &~0 &~0 &~0 &~0 &~0 & -2 & ~2 &~0 & (1,1,1,2,1,2) &~0 &~0 \\ \hline

  \end{tabular}
  \caption{The hidden sectors. All charges are multipled by 4 and the combination $\zeta=1+b_1+b_2+b_3$ has been used.}
  \label{hiddentable}
\end{table}
}

\small{
\begin{table}[h]
\setlength{\tabcolsep}{2.25pt}
\renewcommand{\arraystretch}{1}
  \begin{tabular}{| c | c | c |c| c c c c c c c c | c |  c c |}
    \hline

    F & Sector  & Name & $(C,L)$ & $Q_C$ & $Q_L$ & $Q_{\bar{\eta}^1}$ & $Q_{\bar{\eta}^2}$ & $Q_{\bar{\eta}^3}$ & $Q_{\bar{y}^{3,6}}$ & $Q_{\bar{y}^1\bar{w}^5}$ & $Q_{\bar{w}^{2,4}}$ & $SU(2)_{1,...,6}$ & $Q_{7}$ & $Q_{8}$ \\ \hline

    b & $\alpha+\beta$  &$N_1$& (1,1) &~0 &~0 & -2 & ~2 &~0 &~0 &~0 &~0 & (1,2,1,2,1,1) & ~0 &~0 \\
      &                 &$\bar{N}_1$& (1,1) &~0 &~0 & ~2 & -2 &~0 &~0 &~0 &~0 & (1,2,1,2,1,1) & ~0 &~0 \\
      &                 &$N_2$& (1,1) &~0 &~0 & ~2 & -2 &~0 &~0 &~0 &~0 & (1,2,1,2,1,1) & ~0 &~0 \\
      &                 &$\bar{N}_2$& (1,1) &~0 &~0 & -2 & ~2 &~0 &~0 &~0 &~0 & (1,2,1,2,1,1) & ~0 &~0 \\ \hline
      
    b & $\alpha+\beta+\xi$   &$N_3$& (1,1) &~0 &~0 & -2 & -2 &~0 &~0 &~0 &~0 & (1,1,1,1,1,2) &~4  &~0 \\
      &                      &$\bar{N}_3$& (1,1) &~0 &~0 & ~2 & ~2 &~0 &~0 &~0 &~0 & (1,1,1,1,1,2) & -4 &~0 \\
      &                      &$N_4$& (1,1) &~0 &~0 & -2 & -2 &~0 &~0 &~0 &~0 & (1,1,1,1,1,2) &~4  &~0 \\
      &                      &$\bar{N}_4$& (1,1) &~0 &~0 & ~2 & ~2 &~0 &~0 &~0 &~0 & (1,1,1,1,1,2) & -4 &~0 \\ \hline

\end{tabular}
  \caption{$SO(10)$ singlets without $\tilde{S}$-partners. All charges are multipled by 4 and the combination $\zeta=1+b_1+b_2+b_3$ has been used.}
  \label{nonSmapped1}
\end{table}
}

\small{
\begin{table}[h]
\setlength{\tabcolsep}{2.25pt}
\renewcommand{\arraystretch}{1}
  \begin{tabular}{| c | c | c | c| c c c c c c c c | c |  c c |}
    \hline

    F & Sector &Name& $(C,L)$ & $Q_C$ & $Q_L$ & $Q_{\bar{\eta}^1}$ & $Q_{\bar{\eta}^2}$ & $Q_{\bar{\eta}^3}$ & $Q_{\bar{y}^{3,6}}$ & $Q_{\bar{y}^1\bar{w}^5}$ & $Q_{\bar{w}^{2,4}}$ & $SU(2)_{1,...,6}$ & $Q_{7}$ & $Q_{8}$ \\ \hline

    f & $b_1 +~2\gamma$ &$V_1$& (1,1) &~0 &~0 &~0 & -2 &~2 & -2 &~0 &~0 & (1,1,1,1,1,1) & -4 & -4 \\
      &                 &$V_2$& (1,1) &~0 &~0 &~0 & -2 &~2 & -2 &~0 &~0 & (1,1,1,1,1,1) &~4  &~4  \\
      &                 &$V_3$& (1,1) &~0 &~0 &~0 & -2 &~2 &~2  &~0 &~0 & (1,1,1,1,1,1) & -4 &~4  \\
      &                 &$V_4$& (1,1) &~0 &~0 &~0 & -2 &~2 &~2  &~0 &~0 & (1,1,1,1,1,1) &~4  & -4 \\
      &                 &$V_5$& (1,1) &~0 &~0 &~0 & -2 &~2 & -2  &~0 &0 & (1,1,1,1,2,2) &~0  &~0  \\ \hline
    f & $b_2 +~2\gamma$ &$V_6$& (1,1) &~0 &~0 & -2 &~0 &~2 &~0 & ~2 &~0 & (1,1,1,1,1,1) & -4 & -4 \\
      &                 &$V_7$& (1,1) &~0 &~0 & -2 &~0 &~2 &~0 & ~2 &~0 & (1,1,1,1,1,1) &~4  &~4  \\
      &                 &$V_8$& (1,1) &~0 &~0 & -2 &~0 &~2 &~0 & -2 &~0 & (1,1,1,1,1,1) & -4 &~4  \\
      &                 &$V_9$& (1,1) &~0 &~0 & -2 &~0 &~2 &~0 & -2 &~0 & (1,1,1,1,1,1) &~4  & -4 \\ 
      &                 &$V_{10}$& (1,1) &~0 &~0 & -2 &~0 &~2 &~0 & ~2 &~0 & (1,1,1,1,2,2) &~0  &~0  \\
      \hline
    f & $b_3 +~2\gamma$ &$V_{11}$& (1,1) &~0 &~0 & -2 & -2 &~0 &~0 &~0 & -2 & (1,1,1,1,1,1) & -4 & -4 \\
      &                 &$V_{12}$& (1,1) &~0 &~0 & -2 & -2 &~0 &~0 &~0 & -2 & (1,1,1,1,1,1) & ~4 & ~4 \\
      &                 &$V_{13}$& (1,1) &~0 &~0 & -2 & -2 &~0 &~0 &~0 &~2  & (1,1,1,1,1,1) & -4 &~4  \\
      &                 &$V_{14}$& (1,1) &~0 &~0 & -2 & -2 &~0 &~0 &~0 &~2  & (1,1,1,1,1,1) & ~4 & -4 \\
      &                 &$V_{15}$& (1,1) &~0 &~0 & -2 & -2 &~0 &~0 &~0 & -2 & (1,1,1,1,2,2) & ~0 &~0  \\
      \hline
    f & $b_1+2\gamma+\zeta$ &$V_{16}$& (1,1) &~0 &~0 &~0 & -2 &~2 & ~2 &~0 &~0 & (1,2,2,1,1,1) &~0 &~0 \\
      &                     &$V_{17}$& (1,1) &~0 &~0 &~0 & -2 &~2 & -2 &~0 &~0 & (2,1,1,2,1,1) &~0 &~0 \\ \hline
    f & $b_2+2\gamma+\zeta$ &$V_{18}$& (1,1) &~0 &~0 & -2 &~0 &~2 &~0 & -2 &~0 & (1,2,2,1,1,1) &~0 &~0 \\
      &                     &$V_{19}$& (1,1) &~0 &~0 & -2 &~0 &~2 &~0 & ~2 &~0 & (2,1,1,2,1,1) &~0 &~0 \\ \hline
     
    f & $b_3+2\gamma+\zeta$ &$V_{20}$& (1,1) &~0 &~0 & -2 & -2 &~0 &~0 &~0 & ~2  & (1,2,2,1,1,1) &~0 &~0 \\
      &                     &$V_{21}$& (1,1) &~0 &~0 & -2 & -2 &~0 &~0 &~0 & -2  & (2,1,1,2,1,1) &~0 &~0 \\ \hline

    b & $\tilde{S}+b_1+2\gamma$ &$V_{22}$& (1,1) &~0 &~0  &~0 &~2 &~2  & -2 &~0 &~0 & (1,1,1,2,1,1) &~0 & -4 \\
      &                         &$V_{23}$& (1,1) &~0 &~0  &~0 &~2 &~2  & ~2 &~0 &~0 & (1,1,1,2,1,1) &~0 &~4  \\
      &                         &$V_{24}$& (1,1) &~0 &~0  &~0 &~2 & -2 & ~2 &~0 &~0 & (1,1,2,1,1,2) &~0 &~0  \\ \hline
      
    b & $\tilde{S} + b_2 + 2\gamma$  &$V_{25}$& (1,1) &~0 &~0 & ~2 &~0 & ~2 &~0 & -2  &~0 & (1,1,1,2,1,1) &~0 & ~4 \\
      &                             &$V_{26}$& (1,1) &~0 &~0 & ~2 &~0 & ~2 &~0 & ~2  &~0 & (1,1,1,2,1,1) &~0 & -4 \\ 
      &              &$V_{27}$& (1,1) &~0 &~0 & ~2 &~0 & -2 &~0 & -2  &~0 & (1,1,2,1,1,2) &~0 &~0  \\
      \hline
    
    b & $\tilde{S} + b_3 +~2\gamma$ &$V_{28}$& (1,1) &~0 &~0 & ~2 & -2 &~0 &~0 &~0 & ~2 & (1,1,1,2,1,1) &~0 &~4  \\
      &                             &$V_{29}$& (1,1) &~0 &~0 & ~2 & -2 &~0 &~0 &~0 & -2 & (1,1,1,2,1,1) &~0 & -4 \\
      &                             &$V_{30}$& (1,1) &~0 &~0 & ~2 &~2  &~0 &~0 &~0 & ~2 & (1,1,2,1,1,2) &~0 &~0  \\ \hline

    b & $\tilde{S}+b_1+2\gamma+\zeta$ &$V_{31}$& (1,1) &~0 &~0 &~0 & ~2 & ~2 & ~2 &~0 &~0 & (2,1,1,1,1,1) &   -4 &~0 \\
      &     &$V_{32}$& (1,1) &~0 &~0 &~0 & ~2 & ~2 & -2 &~0 &~0 & (2,1,1,1,1,1) & ~4 &~0 \\
      &                   &$V_{33}$& (1,1) &~0 &~0 &~0 & ~2 & -2 & -2 &~0 &~0 & (1,2,1,1,2,1) & ~0 &~0 \\ \hline

    b & $\tilde{S}+b_2+2\gamma+\zeta$ &$V_{34}$& (1,1) &~0 &~0 & ~2 &~0 & ~2 &~0 & ~2 &~0 & (2,1,1,1,1,1) & ~4 &~0 \\
      &     &$V_{35}$& (1,1) &~0 &~0 & ~2 &~0 & ~2 &~0 & -2 &~0 & (2,1,1,1,1,1) & -4 &~0 \\
      &                   &$V_{36}$& (1,1) &~0 &~0 & ~2 &~0 & -2 &~0 & ~2 &~0 & (1,2,1,1,2,1) & ~0 &~0 \\ \hline

    b & $\tilde{S}+b_3+2\gamma+\zeta$ &$V_{37}$& (1,1) &~0 &~0 & ~2 & -2 &~0 &~0 &~0 & -2 & (2,1,1,1,1,1) & ~4 &~0 \\
      &    &$V_{38}$& (1,1) &~0 &~0 & ~2 & -2 &~0 &~0 &~0 & ~2 & (2,1,1,1,1,1) & -4 &~0 \\
      &                   &$V_{39}$& (1,1) &~0 &~0 & ~2 & ~2 &~0 &~0 &~0 & -2 & (1,2,1,1,2,1) & ~0 &~0 \\ \hline

\end{tabular}
  \caption{$SO(10)$ singlets with $\tilde{S}$-partners. 
All charges are multipled by 4 and the combination $\zeta=1+b_1+b_2+b_3$ 
has been used.}
\label{so10sinws}
\end{table}
}

\small{
\begin{table}[h]
\setlength{\tabcolsep}{2.25pt}
\renewcommand{\arraystretch}{1}
  \begin{tabular}{| c | c | c | c | c c c c c c c c | c | c c |}
    \hline

    F & Sector & Name & $(C,L)$ & $Q_C$ & $Q_L$ & $Q_{\bar{\eta}^1}$ & $Q_{\bar{\eta}^2}$ & $Q_{\bar{\eta}^3}$ & $Q_{\bar{y}^{3,6}}$ & $Q_{\bar{y}^1\bar{w}^5}$ & $Q_{\bar{w}^{2,4}}$ & $SU(2)_{1,...,6}$ & $Q_{7}$ & $Q_{8}$ \\ \hline

    f & $b_2 + \beta$  &$H_1$& (1,2) &~0 &~0 &~0 &~0 &~0 & ~2 &~0 & -2  & (1,1,2,1,1,1) &~0 &~0 \\
      &                &$\bar{H}_1$& (1,2) &~0 &~0 &~0 &~0 &~0 & -2 &~0 & ~2  & (1,1,2,1,1,1) &~0 &~0 \\ \hline

    b & $\tilde{S} + b_2 + \beta$ &$H_2$& (1,2) &~0 &~0 &~0 &~0 &~0 & ~2 &~0 & -2  & (1,1,1,1,2,1) &~0 &~0 \\
      &                           &$\bar{H}_2$& (1,2) &~0 &~0 &~0 &~0 &~0 & -2 &~0 & ~2  & (1,1,1,1,2,1) &~0 &~0 \\ \hline

    b & $b_2 + b_3 $        &$H_3$& (1,1) & -3 & ~2 & ~1 & ~1 & -1 & -2 &~0 &~0 & (1,1,2,1,1,1) & ~2 & ~2 \\
      & $+\beta \pm \gamma$ &$\bar{H}_3$& (1,1) & ~3 & -2 & -1 & -1 & ~1 & ~2 &~0 &~0 & (1,1,2,1,1,1) & -2 & -2 \\ \hline

    f & $\tilde{S}+b_2+b_3$ &$H_4$& (1,1) & -3 & ~2 & ~1 & ~1 & -1 & -2 &~0 &~0 & (1,1,1,1,2,1) & ~2 &~2  \\
      & $+\beta\pm\gamma$   &$H_5$& (1,1) & -3 & ~2 & ~1 & ~1 & -1 & -2 &~0 &~0 & (1,1,1,1,1,2) & -2 & -2 \\
      &                     &$\bar{H}_4$& (1,1) & ~3 & -2 & -1 & -1 & ~1 & ~2 &~0 &~0 & (1,1,1,1,2,1) & -2 & -2  \\
      &                     &$\bar{H}_5$& (1,1) & ~3 & -2 & -1 & -1 & ~1 & ~2 &~0 &~0 & (1,1,1,1,1,2) & ~2 & ~2  \\ \hline

    b & $b_1+b_3$            &$H_6$& (1,1) & -3 & ~2 & ~1 & ~1 & -1 &~0 & -2 &~0 & (1,1,1,2,1,1) & -2 & -2 \\
      & $+\alpha\pm\gamma+\zeta$ &$\bar{H}_6$& (1,1) & ~3 & -2 & -1 & -1 & ~1 &~0 & ~2 &~0 & (1,1,1,2,1,1) & ~2 & ~2 \\ \hline

    f & $\tilde{S}+b_1+b_3$ &$H_7$& ($\bar{3}$,1)         & ~1 & ~2 & ~1 & ~1 & -1 &~0 & ~2 &~0 & (1,1,1,1,1,1) & ~2 & -2 \\
      &$+\alpha\pm\gamma+\zeta$ &$\bar{H}_7$& ($3$,1) & -1 & -2 & -1 & -1 & ~1 &~0 & -2 &~0 & (1,1,1,1,1,1) & -2 & ~2 \\
      &                   &$H_8$& (1,2)         & -3 & -2 & ~1 & ~1 & -1 &~0 & -2 &~0 & (1,1,1,1,1,1) & ~2 & -2 \\
      &                   &$\bar{H}_8$& (1,2)         & ~3 & ~2 & -1 & -1 & ~1 &~0 & ~2 &~0 & (1,1,1,1,1,1) & -2 &  2 \\
      &                   &$H_9$& (1,1)         & -3 & ~2 & -3 & ~1 & -1 &~0 & ~2 &~0 & (1,1,1,1,1,1) & ~2 & -2 \\
      &                   &$\bar{H}_9$& (1,1)         & ~3 & -2 & ~3 & -1 &  1 &~0 & -2 &~0 & (1,1,1,1,1,1) & -2 & ~2 \\
      &                   &$H_{10}$& (1,1)         & -3 & ~2 & ~1 & -3 & -1 &~0 & ~2 &~0 & (1,1,1,1,1,1) & ~2 & -2 \\
      &                   &$\bar{H}_{10}$& (1,1)         & ~3 & -2 & -1 & ~3 & ~1 &~0 & -2 &~0 & (1,1,1,1,1,1) & -2 & ~2 \\
      &                   &$H_{11}$& (1,1)         & -3 & ~2 & ~1 & ~1 & ~3 &~0 & ~2 &~0 & (1,1,1,1,1,1) & ~2 & -2 \\
      &                   &$\bar{H}_{11}$& (1,1)         & ~3 & -2 & -1 & -1 & -3 &~0 & -2 &~0 & (1,1,1,1,1,1) & -2 & ~2 \\ \hline
      
      f & $b_3 \pm \gamma$ &$H_{12}$& ($\bar{3}$,1)         &~1  & -2 & ~1 & ~1 & -1 &~0 &~0 & -2 & (1,1,1,1,1,1) &~2  &~2  \\
      &                    &$\bar{H}_{12}$& ($3$,1) & -1 & ~2 & -1 & -1 & ~1 &~0 &~0 &~2  & (1,1,1,1,1,1) & -2 & -2 \\
      &                    &$H_{13}$& (1,2)         & -3 & ~2 & ~1 & ~1 & -1 &~0 &~0 & -2 & (1,1,1,1,1,1) &~2  &~2  \\
      &                    &$\bar{H}_{13}$& (1,$2$) & ~3 & -2 & -1 & -1 & ~1 &~0 &~0 &~2  & (1,1,1,1,1,1) & -2 & -2 \\
      &                    &$H_{14}$& (1,1)         & -3 & -2 & -3 & ~1 & -1 &~0 &~0 & -2 & (1,1,1,1,1,1) &~2  &~2  \\
      &                    &$\bar{H}_{14}$& (1,1)         & ~3 & ~2 &~3  & -1 & ~1 &~0 &~0 &~2  & (1,1,1,1,1,1) & -2 & -2 \\
      &                    &$H_{15}$& (1,1)         & -3 & -2 & ~1 & -3 & -1 &~0 &~0 & -2 & (1,1,1,1,1,1) &~2  &~2  \\
      &                    &$\bar{H}_{15}$& (1,1)         & ~3 & ~2 & -1 &~3  & ~1 &~0 &~0 &~2  & (1,1,1,1,1,1) & -2 & -2 \\
      
       &                    &$H_{16}$& (1,1)         & -3 & -2 & ~1 & ~1 &~3  &~0 &~0 & -2 & (1,1,1,1,1,1) &~2  &~2  \\
      &                    &$\bar{H}_{16}$& (1,1)         & ~3 & ~2 & -1 & -1 & -3 &~0 &~0 &~2  & (1,1,1,1,1,1) & -2 & -2 \\ \hline
      
    b & $\tilde{S} + b_3 \pm \gamma$ &$H_{17}$& (1,1) & -3 & -2 & ~1 & ~1 & -1 &~0 &~0 &  2 & (1,1,1,2,1,1) & -2 & ~2 \\
      &                              &$\bar{H}_{17}$& (1,1) & ~3 & ~2 & -1 & -1 & ~1 &~0 &~0 & -2 & (1,1,1,2,1,1) & ~2 & -2 \\ \hline

    b & $b_1+b_2$    &$H_{18}$& (1,1) & ~3 & ~2 & ~1 & ~1 & ~1 & ~2 & -2 &~0 & (1,1,1,1,2,1) & ~2 & -2 \\
      & $\alpha+\beta\pm\gamma+\zeta$ &$H_{19}$& (1,1) & ~3 & ~2 & ~1 & ~1 & ~1 & ~2 & ~2 &~0 & (1,1,1,1,1,2) & -2 & ~2 \\
      &                   &$H_{20}$& (1,1) & -3 & -2 & -1 & -1 & -1 & ~2 & -2 &~0 & (1,1,1,1,2,1) & -2 &~2  \\
      &                   &$H_{21}$& (1,1) & -3 & -2 & -1 & -1 & -1 & ~2 & ~2 &~0 & (1,1,1,1,1,2) & ~2 & -2 \\ \hline 
    f & $\tilde{S}+b_1+b_2$        &$H_{22}$& (1,1) & ~3 & ~2 & ~1 & ~1 & ~1 & -2 & ~2 &~0 & (1,1,2,1,1,1) & ~2 & -2 \\
      & $+\alpha+\beta\pm\gamma+\zeta$ &$H_{23}$& (1,1) & -3 & -2 & -1 & -1 & -1 & -2 & ~2 &~0 & (1,1,2,1,1,1) & -2 & ~2 \\ \hline
\end{tabular}
  \caption{Exotic states with $\tilde{S}$-partners (i). All charges are multipled by 4 and the combination $\zeta=1+b_1+b_2+b_3$ has been used.}
\label{exotics1}
\end{table}
}

\small{
\begin{table}[h]
\setlength{\tabcolsep}{2.25pt}
\renewcommand{\arraystretch}{1}
  \begin{tabular}{| c | c | c | c | c c c c c c c c | c | c c |}
    \hline

    F & Sector & Name & $(C,L)$ & $Q_C$ & $Q_L$ & $Q_{\bar{\eta}^1}$ & $Q_{\bar{\eta}^2}$ & $Q_{\bar{\eta}^3}$ & $Q_{\bar{y}^{3,6}}$ & $Q_{\bar{y}^1\bar{w}^5}$ & $Q_{\bar{w}^{2,4}}$ & $SU(2)_{1,...,6}$ & $Q_{7}$ & $Q_{8}$ \\ \hline

    b & $b_1+b_3$    &$H_{24}$& (1,1) & ~3 & ~2 & ~1 & -1 & -1 & ~2 &~0 & ~2 & (1,1,1,1,2,1) &~2  & -2 \\
      & $+\alpha+\beta\pm\gamma+\zeta$ &$H_{25}$& (1,1) & ~3 & ~2 & ~1 & -1 & -1 & ~2 &~0 & -2 & (1,1,1,1,1,2) & -2 &~2  \\
      &                   &$H_{26}$& (1,1) & -3 & -2 & -1 & ~1 & ~1 & ~2 &~0 & ~2 & (1,1,1,1,2,1) & -2 &~2  \\
      &                   &$H_{27}$& (1,1) & -3 & -2 & -1 & ~1 & ~1 & ~2 &~0 & -2 & (1,1,1,1,1,2) &~2  & -2 \\ \hline

    f & $\tilde{S}+b_1+b_3$        &$H_{28}$& (1,1) & ~3 & ~2 & ~1 & -1 & -1 & -2 &~0 & -2 & (1,1,2,1,1,1) &  2 &-2  \\
      & $+\alpha+\beta\pm\gamma+\zeta$ &$H_{29}$& (1,1) & -3 & -2 & -1 & ~1 & ~1 & -2 &~0 & -2 & (1,1,2,1,1,1) & -2 &  2 \\ \hline

    b & $b_2+b_3$    &$H_{30}$& (1,1) & ~3 & ~2 & -1 & ~1 & -1 &~0 & -2 & ~2 & (1,1,1,1,2,1) & ~2 & -2 \\
      & $+\alpha+\beta\pm\gamma+\zeta$ &$H_{31}$& (1,1) & ~3 & ~2 & -1 & ~1 & -1 &~0 & -2 & -2 & (1,1,1,1,1,2) & -2  &~2 \\
      &                   &$H_{32}$& (1,1) & -3 & -2 & ~1 & -1 & ~1 &~0 & -2 & ~2 & (1,1,1,1,2,1) & -2 & ~2 \\
      &                   &$H_{33}$& (1,1) & -3 & -2 & ~1 & -1 & ~1 &~0 & -2 & -2 & (1,1,1,1,1,2) & ~2 & -2 \\ \hline

    f & $\tilde{S}+b_2+b_3$        &$H_{34}$& (1,1) & ~3 & ~2 & -1 & ~1 & -1 &~0 & ~2 & -2 & (1,1,2,1,1,1) &~2 & -2 \\
      & $+\alpha+\beta\pm\gamma+\zeta$ &$H_{35}$& (1,1) & -3 & -2 & ~1 & -1 & ~1 &~0 & ~2 & -2 & (1,1,2,1,1,1) & -2 &~2 \\ \hline

      \end{tabular}
  \caption{Exotic states with $\tilde{S}$-partners (ii). All charges are multipled by 4 and the combination $\zeta=1+b_1+b_2+b_3$ has been used.}
\label{exotics2}
\end{table}
}

\small{
\begin{table}[h]
\setlength{\tabcolsep}{2.25pt}
\renewcommand{\arraystretch}{1}
  \begin{tabular}{| c | c | c | c | c c c c c c c c | c | c c |}
    \hline

    F & Sector  &Name& $(C,L)$ & $Q_C$ & $Q_L$ & $Q_{\bar{\eta}^1}$ & $Q_{\bar{\eta}^2}$ & $Q_{\bar{\eta}^3}$ & $Q_{\bar{y}^{3,6}}$ & $Q_{\bar{y}^1\bar{w}^5}$ & $Q_{\bar{w}^{2,4}}$ & $SU(2)_{1,...,6}$ & $Q_{7}$ & $Q_{8}$ \\ \hline
    
    f & $b_1 + \alpha$ &$H_{36}$& (1,$2$) &~0 &~0 &~0 &~0 &~0 &~0 & -2 & ~2 & (2,1,1,1,1,1) &~0 &~0 \\
      &                &$\bar{H}_{36}$& (1,$2$) &~0 &~0 &~0 &~0 &~0 &~0 & ~2 & -2 & (2,1,1,1,1,1) &~0 &~0 \\ \hline
        
    b & $b_1+b_3$             &$H_{37}$& (1,1) & -3 & ~2 & ~1 & ~1 & -1 &~0 & -2 &~0 & (2,1,1,1,1,1) & ~2 &~2 \\
      & $+\alpha \pm \gamma$  &$\bar{H}_{37}$& (1,1) & ~3 & -2 & -1 & -1 & ~1 &~0 & ~2 &~0 & (2,1,1,1,1,1) & -2 &-2 \\ \hline
    b & $\tilde{S}+b_3$  &$H_{38}$& (1,1) & -3 & -2 & ~1 & ~1 & -1 &~0 &~0 & -2 & (2,1,1,1,1,1) &~2  & -2 \\
      &  $\pm\gamma+\zeta$   &$\bar{H}_{38}$& (1,1) & ~3 & ~2 & -1 & -1 & ~1 &~0 &~0 & ~2 & (2,1,1,1,1,1) & -2 &~2  \\ \hline

    b & $b_2+b_3$            &$H_{39}$& (1,1) & -3 & ~2 & ~1 & ~1 & -1 & ~2 &~0 &~0 & (1,2,1,1,1,1) & -2 & -2 \\
      & $+\beta\pm\gamma+\zeta$  &$\bar{H}_{39}$& (1,1) & ~3 & -2 & -1 & -1 & ~1 & -2 &~0 &~0 & (1,2,1,1,1,1) & ~2 & ~2 \\ \hline

    f & $\tilde{S}+b_2+b_3$       &$H_{40}$& (1,1) & ~3 & ~2 & -1 &  1 & -1 &~0 & ~2 & -2 & (1,2,1,1,1,1) &  -2 & ~2 \\
      & $+\alpha+\beta\pm\gamma$  &$H_{41}$& (1,1) & -3 & -2 &  1 & -1 & ~1 &~0 & ~2 & -2 & (1,2,1,1,1,1) &  ~2 & -2 \\ \hline

    f & $\tilde{S}+b_1+b_3$       &$H_{42}$& (1,1) & ~3 & ~2 & -1 & ~1 & -1 & -2 &~0 & -2 & (1,2,1,1,1,1) & -2 &~2  \\
      & $+\alpha+\beta\pm\gamma$  &$H_{43}$& (1,1) & -3 & -2 &  1 & -1 & ~1 & -2 &~0 & -2 & (1,2,1,1,1,1) & ~2 & -2 \\ \hline

    f & $\tilde{S}+b_1+b_2$      &$H_{44}$& (1,1) & ~3 & ~2 & ~1 & ~1 & -1 & -2 &~2 &~0 & (1,2,1,1,1,1) &  -2 &~2  \\
      & $+\alpha+\beta\pm\gamma$ &$H_{45}$&  (1,1) & -3 & -2 & -1 & -1 & ~1 & -2 &~2 &~0 & (1,2,1,1,1,1) &  ~2 & -2 \\ \hline

      \end{tabular}
  \caption{Exotic spinorials without $\tilde{S}$-partners (i). All charges are multipled by 4 and the combination $\zeta=1+b_1+b_2+b_3$ has been used.}
  \label{NonSmapped3}
\end{table}
}

\small{
\begin{table}[h]
\setlength{\tabcolsep}{2.25pt}
  \begin{tabular}{| c | c  | c | c | c c c c c c c c | c | c c |}
    \hline

    F & Sector & Name & $(C,L)$ & $Q_C$ & $Q_L$ & $Q_{\bar{\eta}^1}$ & $Q_{\bar{\eta}^2}$ & $Q_{\bar{\eta}^3}$ & $Q_{\bar{y}^{3,6}}$ & $Q_{\bar{y}^1\bar{w}^5}$ & $Q_{\bar{w}^{2,4}}$ & $SU(2)_{1,...,6}$ & $Q_{7}$ & $Q_{8}$ \\ \hline
    b & $\tilde{S}+b_1+b_2+b_3$  &$H_{46}$& (1,2) &~0 & ~0 &~0 & -2 & ~2 &~0 &~0 &~0 & (1,1,1,1,1,1) &~0 & -4 \\
    &       $+\alpha+\zeta$     &$\bar{H}_{46}$& (1,2) &~0 & ~0 &~0 & ~2 & -2 &~0 &~0 &~0 & (1,1,1,1,1,1) &~0 & ~4 \\
    
    &                       &$H_{47}$& (1,2) &~0 & ~0 &~0 & -2 & ~2 &~0 &~0 &~0 & (1,1,1,1,1,1) &~0 & -4 \\
    &                       &$\bar{H}_{47}$& (1,2) &~0 & ~0 &~0 & ~2 & -2 &~0 &~0 &~0 & (1,1,1,1,1,1) &~0 & ~4 \\
      &                       &$H_{48}$& (1,1) &~0 & ~4 &~0 & -2 & -2 &~0 &~0 &~0 & (1,1,1,1,1,1) &~0 & -4 \\
      &                       &$\bar{H}_{48}$& (1,1) &~0 & -4 &~0 & ~2 & ~2 &~0 &~0 &~0 & (1,1,1,1,1,1) &~0 & ~4 \\
      
      &                       &$H_{49}$& (1,1) &~0 & ~4 &~0 & ~2 & ~2 &~0 &~0 &~0 & (1,1,1,1,1,1) &~0 & -4 \\
      &                       &$\bar{H}_{49}$& (1,1) &~0 & -4 &~0 & -2 & -2 &~0 &~0 &~0 & (1,1,1,1,1,1) &~0 & ~4 \\
      &                       &$H_{50}$& (1,1) &~0 & ~4 &~0 & ~2 & ~2 &~0 &~0 &~0 & (1,1,1,1,1,1) &~0 & -4 \\
      &                       &$\bar{H}_{50}$& (1,1) &~0 & -4 &~0 & -2 & -2 &~0 &~0 &~0 & (1,1,1,1,1,1) &~0 & ~4 \\
      &                       &$H_{51}$& (1,1) &~0 & ~4 &~0 & -2 & -2 &~0 &~0 &~0 & (1,1,1,1,1,1) &~0 & -4 \\
      &                       &$\bar{H}_{51}$& (1,1) &~0 & -4 &~0 & ~2 & ~2 &~0 &~0 &~0 & (1,1,1,1,1,1) &~0 & ~4 \\

       \hline

    b & $\tilde{S}+b_1+b_2+b_3$ &$H_{52}$& (3,1)         &~2  &~0 & -2 &~0  &~0 &~0 &~0 &~0 & (1,1,1,1,1,1) &~4  &~0 \\
    &   $+\alpha+2\gamma+\zeta$ &$\bar{H}_{52}$& ($\bar{3}$,1) & -2 &~0 &~2  &~0  &~0 &~0 &~0 &~0 & (1,1,1,1,1,1) & -4 &~0 \\
      &                          &$H_{53}$& (3,1)         &~2  &~0 & -2 &~0  &~0 &~0 &~0 &~0 & (1,1,1,1,1,1) &~4  &~0 \\
      
      &                       &$\bar{H}_{53}$& ($\bar{3}$,1) & -2 &~0 &~2  &~0  &~0 &~0 &~0 &~0 & (1,1,1,1,1,1) & -4 &~0 \\
      &                       &$H_{54}$& (1,1)         &~6  &~0 &~2  &~0  &~0 &~0 &~0 &~0 & (1,1,1,1,1,1) &~4  &~0 \\
      &                      &$\bar{H}_{54}$& (1,1)         & -6 &~0 & -2 &~0  &~0 &~0 &~0 &~0 & (1,1,1,1,1,1) & -4 &~0 \\
      &                       &$H_{55}$& (1,1)         &~6  &~0 &~2  &~0  &~0 &~0 &~0 &~0 & (1,1,1,1,1,1) &~4  &~0 \\
      &                      &$\bar{H}_{55}$& (1,1)         & -6 &~0 & -2 &~0  &~0 &~0 &~0 &~0 & (1,1,1,1,1,1) & -4 &~0 \\
       \hline

    b & $\tilde{S}+b_1+b_2$ &$H_{56}$& (1,2) &~0 &~0  & -2 &~0 & -2 &~0 &~0 &~0 & (1,1,1,1,2,1) &~0 &~0 \\
    &                     &$\bar{H}_{56}$& (1,2) &~0 &~0  &~2  &~0 &~2  &~0 &~0 &~0 & (1,1,1,1,2,1) &~0 &~0 \\
      &     $+b_3+\beta$    &$H_{57}$& (1,2) &~0 &~0  & -2 &~0 & -2 &~0 &~0 &~0 & (1,1,1,1,2,1) &~0 &~0 \\
      
      &                      &$\bar{H}_{57}$& (1,2) &~0 &~0  &~2  &~0 &~2  &~0 &~0 &~0 & (1,1,1,1,2,1) &~0 &~0 \\
      &                      &$H_{58}$& (1,1) &~0 &~4  &~2  &~0 & -2 &~0 &~0 &~0 & (1,1,1,1,2,1) &~0 &~0 \\
      &                      &$\bar{H}_{58}$& (1,1) &~0 & -4 & -2 &~0 &~2  &~0 &~0 &~0 & (1,1,1,1,2,1) &~0 &~0 \\
      &                      &$H_{59}$& (1,1) &~0 &~4  &~2  &~0 & -2 &~0 &~0 &~0 & (1,1,1,1,2,1) &~0 &~0 \\
      &                     &$\bar{H}_{59}$& (1,1) &~0 & -4 & -2 &~0 &~2  &~0 &~0 &~0 & (1,1,1,1,2,1) &~0 &~0 \\

       \hline
       & $\tilde{S}+b_1$      &$H_{60}$& (1,1) &~0 &~4  &~0 &~0 &~0 &~0 & -2 &~2  & (1,1,1,1,1,1) &~0 &~4  \\
    b & $+\alpha+\zeta$  &$\bar{H}_{60}$& (1,1) &~0 & -4 &~0 &~0 &~0 &~0 & -2 &~2  & (1,1,1,1,1,1) &~0 & -4 \\
      
      &                    &$H_{61}$& (1,1) &~0 &~4  &~0 &~0 &~0 &~0 &~2  &~2  & (1,1,1,1,1,1) &~0 & -4 \\
      &                    &$\bar{H}_{61}$& (1,1) &~0 & -4 &~0 &~0 &~0 &~0 &~2  &~2  & (1,1,1,1,1,1) &~0 &~4  \\
      &                    &$H_{62}$& (1,1) &~0 &~4  &~0 &~0 &~0 &~0 &~2  & -2 & (1,1,1,1,1,1) &~0 &~4  \\
      &                    &$\bar{H}_{62}$& (1,1) &~0 & -4 &~0 &~0 &~0 &~0 &~2  & -2 & (1,1,1,1,1,1) &~0 & -4 \\
      
      &                    &$H_{63}$& (1,1) &~0 &~4  &~0 &~0 &~0 &~0 & -2 & -2 & (1,1,1,1,1,1) &~0 & -4 \\
      &                    &$\bar{H}_{63}$& (1,1) &~0 & -4 &~0 &~0 &~0 &~0 & -2 & -2 & (1,1,1,1,1,1) &~0 &~4  \\
       \hline

      \end{tabular}
  \caption{Exotic spinorials without $\tilde{S}$-partners (ii). All charges are multipled by 4 and the combination $\zeta=1+b_1+b_2+b_3$ has been used.}
  \label{nonSmapped4}
\end{table}
}

\end{appendix}
\clearpage

\bibliographystyle{unsrt}

\begin{thebibliography}{99}

\bibitem{fff}
I. Antoniadis, C. Bachas, and C. Kounnas, \NPB{289}{1987}{87};\\
H. Kawai, D.C. Lewellen, and S.H.-H. Tye, \NPB{288}{1987}{1};\\
I. Antoniadis and C. Bachas, \NPB{298}{1988}{586}.

\bibitem{fsu5} I. Antoniadis, J. Ellis, J. Hagelin and D.V. Nanopoulos,
            \PLB{231}{1989}{65}

\bibitem{fny} A.E. Faraggi, D.V. Nanopoulos and K. Yuan,
                    \NPB{335}{1990}{347};\\
              \AEF, \PRD{46}{1992}{3204};\\
              G.B. Cleaver, A.E. Faraggi and D.V. Nanopoulos,
            \PLB{455}{1999}{135}.

\bibitem{alr} I. Antoniadis. G.K. Leontaris and J. Rizos,
                                \PLB{245}{1990}{161};\\
              G.K. Leontaris and J. Rizos, \NPB{554}{1999}{3}.

\bibitem{slm} 
              A.E. Faraggi, \PLB{278}{1992}{131}; \NPB{387}{1992}{239};\\
            \AEF, E. Manno and C.M. Timirgaziu, \EJP{50}{2007}{701}.

\bibitem{lrs}
    G.B. Cleaver, A.E. Faraggi and C. Savage, \PRD{63}{2001}{066001};\\
    G.B. Cleaver, D.J Clements and A.E. Faraggi, \PRD{65}{2002}{106003};

\bibitem{acfkr}
B. Assel, C. Christodoulides, A.E. Faraggi, C. Kounnas and J. Rizos
                             \PLB{683}{2010}{306}; \NPB{844}{2011}{365}; \\
 C. Christodoulides, A.E. Faraggi and J. Rizos,
                             \PLB{702}{2011}{81}.

\bibitem{su62} L. Bernard \etal, \NPB{868}{2013}{1}.

\bibitem{frs}
	\AEF, J. Rizos and H. Sonmez, \NPB{886}{2014}{202}; \\
        H. Sonmez, \PRD{93}{2016}{125002}.

\bibitem{slmclass} \AEF, J. Rizos and H. Sonmez, \NPB{927}{2018}{1}.

\bibitem{lrsclass} \AEF, G. Harries and J. Rizos, \NPB{936}{2018}{472}. 

\bibitem{coscos} {\it See e.g.:} C. Kounnas and N. Toumbas, 
 {\it PoS\ Corfu\ } {\bf 2012} (2013) 083, arXiv:1305.2809, 
and references therein. 

\bibitem{z2xz2} \AEF, \PLB{326}{1994}{62};\\
                E. Kiritsis and C. Kounnas, \NPB{503}{1997}{117};\\
                \AEF, S. Forste and C. Timirgaziu, \JHEP{0608}{2006}{057};\\
                P. Athanasopoulos, \AEF, S. Groot Nibbelink and V.M. Mehta,
                                          \JHEP{1604}{2016}{038}.

\bibitem{tw} {\it See e.g.}: A. Taormina and K. Wendland, arXiv:1908.03148. 

\bibitem{dh} L.J. Dixon, J.A. Harvey, \NPB{274}{1986}{93};\\ 
 L. Alvarez--Gaume, P.H. Ginsparg, G.W. Moore and C. Vafa, \PLB{171}{1986}{155}.

\bibitem{gv} P.H. Ginsparg and C. Vafa, \NPB{289}{1986}{414}.

\bibitem{itoyama} H. Itoyama and T.R. Taylor, \PLB{186}{1987}{129}. 

\bibitem{kltclas} H. Kawai, D.C. Lewellen and S.H.H. Tye, \PRD{34}{1986}{3794}.

\bibitem{nonsusy} K.R. Dienes, \PRL{65}{1990}{1979}; \PRD{42}{1990}{2004};\\
  S. Abel and K.R. Dienes, \PRD{91}{2015}{126014};\\
  M.~Blaszczyk, S.~Groot Nibbelink, O.~Loukas and F.~Ruehle,
                            \JHEP{1510}{2015}{166};\\
  S.~Groot Nibbelink and E.~Parr, \PRD{94}{2016}{041704};\\
  I.~Florakis and J.~Rizos, \NPB{913}{2016}{495};\\
  I.~Florakis, arXiv:1611.10323;\\
  S.~Groot Nibbelink \etal, arXiv:1710.09237;\\
  S. Abel, K.R. Dienes and E. Mavroudi, \PRD{97}{2018}{126017};\\
  T. Coudarchet and H. Partouche, \NPB{933}{2018}{134};\\
  A. Abel, E. Dudas, D. Lewis and H. Partouche, arXiv:1812.09714;\\
  H. Itoyama and S. Nakajima, arXiv:1905.10745;\\
  M. McGuigan, arXiv:1907.01944.
  
\bibitem{interpol} \AEF~and M. Tsulaia, \PLB{683}{2010}{314};\\
  B.~Aaronson, S.~Abel and E.~Mavroudi, \PRD{95}{2017}{106001}. 

\bibitem{aafs}
J.M. Ashfaque, P. Athanasopoulos, \AEF~and H.Sonmez, \EJP{76}{2016}{208}.

\bibitem{spwsp} \AEF, \EJP{79}{2019}{703}.

\bibitem{msds} C. Kounnas, {\it Fortsch.\ Phys.\ } {\bf56} (2008) 1143;\\
  I. Florakis and C. Kounnas, \NPB{820}{2009}{237};\\
  I. Florakis, C. Kounnas and N. Toumbas, \NPB{834}{2010}{273}.

\bibitem{racetrack}
  N.V. Krasnikov, \PLB{193}{1987}{37};\\
  L.J. Dixon, Supersymmetry Breaking in String Theory, 
  in The Rice Meeting: Proceedings, B. Bonner and H. Miettinen, eds.,
  World Scientific (Singapore) 1990.

\bibitem{cfmt} G.B. Cleaver, \AEF, E. Manno and C. Timirgaziu, 
                      \PRD{78}{2008}{046009}.

\bibitem{su421} G.B. Cleaver, A.E. Faraggi and S.E.M. Nooij,
                                                \NPB{672}{2003}{64};\\
                \AEF~and H. Sonmez, \PRD{91}{2015}{066006}.

\bibitem{gkr} A. Gregori, C.~Kounnas and J.~Rizos, \NPB{549}{1999}{16}.

\bibitem{fknr}
	\AEF, C. Kounnas, S.E.M Nooij and J. Rizos, hep-th/0311058;
                                                    \NPB{695}{2004}{41}.
\bibitem{fkr}
	\AEF, C. Kounnas and J. Rizos, \PLB{648}{2007}{84};
                                      \NPB{774}{2007}{208};
                                      \NPB{799}{2008}{19}.

\bibitem{cfkr} T. Catelin-Julian, A.E. Faraggi, C. Kounnas and J. Rizos,
            \NPB{812}{2009}{103};\\
            C. Angelantonj, \AEF~and M. Tsulaia, \JHEP{1007}{2010}{4}.

\bibitem{nahe} \AEF~and D.V. Nanopoulos, \PRD{48}{1993}{3288};\\
               \AEF, \IJMP{14}{1999}{1663}. 

\bibitem{dtsm} \AEF, \NPB{428}{1994}{111}; \PLB{520}{2001}{337}.

\bibitem{topyuk} \AEF, \PLB{274}{1992}{47}; \PRD{47}{1993}{5021}. 
  
\bibitem{moduli} \AEF, \NPB{728}{2005}{83}.

\bibitem{xmap}
	A.E. Faraggi, \NPB{407}{1993}{57}; \EJP{49}{2007}{803}.

\bibitem{ffmt} 
        \AEF, I. Florakis, T. Mohaupt and M. Tsulaia, \NPB{848}{2011}{332}.

\bibitem{cfua1}	G. Cleaver and \AEF, \IJMP{14}{1999}{2335}.

\bibitem{gs} M.B. Green and J.H. Schwarz, \PLB{149}{1984}{117}.

\bibitem{ads} J.J.~Atick, L.J.~Dixon and A.~Sen, \NPB{292}{1987}{109}.

\bibitem{dkl} L.J. Dixon, V. Kaplunovsky and J. Louis, \NPB{329}{1990}{27}. 

\bibitem{stefan} S. Groot Nibbelink, M. Trapletti and M. Walter, 
                       \PLB{652}{2007}{124};\\
                S. Groot Nibbelink, T.W. Ha and M. Trapletti,
                     \PRD{77}{2008}{026002};\\
                S. Groot Nibbelink, D. Klevers, F. Ploger, M. Trapletti
                and P.K.S. Vaundrevange, \JHEP{0804}{2008}{60}. 
           
\bibitem{thirring}
	J. Bagger, D. Nemeschansky, N. Seiberg and
	S. Yankielowicz, \NPB{289}{1987}{53},\\
	D. Chang and A. Kumar,  \PRD{38}{1988}{1893}; \PRD{38}{1988}{3734}.

\bibitem{dsw} M. Dine, N. Seiberg and E. Witten, \NPB{289}{1987}{589}. 

\bibitem{cleaveretal} G. Cleaver \etal, \EJP{71}{2011}{1842}. 



\end{thebibliography}

\end{document}